\documentclass[11pt]{article}
\usepackage{amsfonts,slashed}
\usepackage{arydshln}
\usepackage[greek,english]{babel}
\usepackage{cancel}
\usepackage{upgreek}
\usepackage{float}
\usepackage{url}
\usepackage{latexsym}
\usepackage{amsfonts}
\usepackage{amsmath}
\usepackage{epsfig}
\usepackage{latexsym,amssymb}
\usepackage{amsmath,amssymb,amsthm}
\usepackage{hyperref}
\usepackage{fontenc}
\usepackage{graphicx,tikz}
\usepackage{multirow}
\usepackage{xfrac}
\usepackage[T1]{fontenc}
\usepackage{textcomp}
\usepackage{lmodern}
\usepackage{stmaryrd}
\usepackage{cite}
\usepackage{mathtools}

\usepackage[margin=20pt,small]{caption}
\usepackage{subcaption}

\usepackage{ifpdf}
\ifx\pdfoutput\undefined
   \pdffalse
   \usepackage{cite}
 \else
   \pdfoutput=1
   \pdftrue
\fi
\DeclareFontFamily{OMS}{rsfs}{\skewchar\font'60}
\DeclareFontShape{OMS}{rsfs}{m}{n}{<-5>rsfs5 <5-7>rsfs7 <7->rsfs10 }{}
\DeclareSymbolFont{rsfs}{OMS}{rsfs}{m}{n}
\DeclareSymbolFontAlphabet{\Scr}{rsfs}

\numberwithin{equation}{section}
\def\be{\begin{equation}}
\def\ee{\end{equation}}
\def\ba{\begin{array}}
\def\ea{\end{array}}

\newcommand{\bea}{\begin{eqnarray}}
\newcommand{\eea}{\end{eqnarray}}

\def\={~=~}
\def\*{{}^*}

\newcommand{\ov }[1]{\overline{#1}}

\newcommand{\del}{\partial}
\newcommand{\PR}[1]{(\mathbb{P}_{#1})}
\newcommand{\rK}{{\rm K}}
\newcommand{\rL}{{\rm L}}

\definecolor{darkred}{rgb}{0.65,0.15,0}

\def\={~=~}
\def\*{{}^*}

\newcommand{\SL}[1]{\mathrm{SL}(#1)}
\newcommand{\SU}[1]{\mathrm{SU}(#1)}
\newcommand{\USp}[1]{\mathrm{USp}(#1)}
\newcommand{\SO}[1]{\mathrm{SO}(#1)}
\newcommand{\En}[1]{\mathrm{E}_{#1(#1)}}

\newcommand{\cY}{{\cal Y}}

\newcommand{\cP}{{\cal P}}

\begin{document}

\begin{titlepage}
\vfill
\begin{flushright}

\end{flushright}

\vfill
\begin{center}
	{\LARGE \bf Consistent truncations and  G$_2$-invariant\\[1ex] 
	  AdS$_4$ solutions of $D=11$ supergravity
	}\\[1cm]
	
{\bf Bastien Duboeuf\,$^{a}{\!}$
		\footnote{\tt bastien.duboeuf@ens-lyon.fr}, Michele Galli\,${}^{b}{\!}$
		\footnote{\tt m.galli@uq.edu.au}, Emanuel Malek\,${}^{c}{\!}$
		\footnote{\tt emanuel.malek@gmail.com}, Henning Samtleben\,${}^{a,d}{\!}$
		\footnote{\tt henning.samtleben@ens-lyon.fr}
		}
\vskip .8cm
	
	{\it ${}^a$ ENS de Lyon, CNRS, LPENSL, UMR5672, 69342, Lyon cedex 07, France}\\[2ex]
	{\it  $^{b}$ School of Mathematics and Physics, University of Queensland, \\
	St Lucia, Brisbane, Queensland 4072, Australia}\\[2ex]
	{\it  $^{c}$ Department of Physics, Astronomy and Mathematics, University of Hertfordshire, \\
	College Lane, Hatfield, AL10 9AB, United Kingdom}\\[2ex] 
		{\it  $^{d}$ Institut Universitaire de France (IUF)}
	
\end{center}
%%%%%%%%%%%%%%%%%%%%%%%%%%%%%%%%%%%%%%%%%%%%%%%
\vfill

\begin{center}
	\textbf{Abstract}
	
\end{center}
\begin{quote}
Maximal supergravities in ten and eleven dimensions admit consistent truncations
on particular spheres to maximal supergravities in lower dimensions.
Concurrently, the truncation to singlets under any subgroup of the sphere isometry group
leads to consistent truncations with less or no supersymmetry. We review the relation between
these truncations in the framework of exceptional field theory. As an application, we derive three
new G$_2$-invariant solutions of $D=11$ supergravity. Their geometry is of the form AdS$_4\times \Sigma_7$
where $\Sigma_7$ is a deformed seven-sphere, preserving SO(7) isometries.

\end{quote}
\vfill
\setcounter{footnote}{0}

\end{titlepage}

\tableofcontents \noindent {}

\newpage

\section{Introduction}

Generalised geometry and exceptional field theory (ExFT) have proven to be invaluable tools in the construction of consistent truncations of type II and 11-dimensional supergravity \cite{Lee:2014mla,Hohm:2014qga,Cassani:2019vcl}. In particular, the language of generalised $G$-structures reduces the problem of constructing a truncation ansatz, to that of understanding the generalised intrinsic torsion of a particular generalised $G$-structure \cite{Cassani:2019vcl}. An interesting example are maximally supersymmetric truncations on spheres or products thereof, which arise from the generalised Leibniz parallelisability of spheres $S^n$ \cite{Lee:2014mla,Hohm:2014qga}. These truncations are powerful because they always contain a large number of scalar fields, which is a promising starting point when one is looking for new AdS$_d$ solutions. 
Specifically, the scalars in maximal supergravity in $D=11-d$ dimensions parametrise the target space
\begin{equation}
{\cal M}_{\rm scalar}= \frac {{\rm E}_{d(d)}}{{\rm K}_{d}}\,,
\label{eq:cosetEK}
\end{equation}
where ${\rm K}_{d}$ denotes the maximal compact subgroup of the exceptional group ${\rm E}_{d(d)}$.
More recently, exceptional field theory has also been adopted as a universal tool to compute the full Kaluza-Klein (KK) spectra around any solution which fits within a maximally supersymmetric truncation \cite{Malek:2019eaz,Malek:2020yue,Cesaro:2020soq}.

The key advantage of using ExFT for the computation of KK masses, is that it allows one to express all fluctuations in terms of only the scalar harmonics on the internal manifold, without having to resort to any tensorial or spinorial harmonics. This is because all the non-trivial tensor structure is encoded in the generalised frame that defines the parallelisation. This feature is retained whenever the internal space is generalised parallelisable, i.e.\ crucially the backgrounds that one considers do not necessarily have to fit within a maximal truncation. This observation was used in \cite{Duboeuf:2022mam,Duboeuf:2023dmq} to compute the spectra around the supersymmetric AdS$_4\times S^7_{\rm squashed}$ background, which is not contained within maximal ${\cal N}=8$ supergravity.

A complementary approach to consistent truncations which has been exploited since the early years of supergravity \cite{Duff:1985jd} makes use of the fact that the truncation of a higher-dimensional theory to all fields invariant under a subgroup ${\rm K}$ of the isometry group ${\rm SO}(n+1)$ of the internal space $M_{\rm int}=S^n$ is automatically consistent. This is because the retained ${\rm K}$-singlet fields cannot source the truncated non-singlet fields. Consequently, consistency in general requires to retain all the ${\rm K}$-singlet fields. In general, such a truncation will contain an infinite number of fields (including in particular an infinite number of massive spin-2 fluctuations), unless the group ${\rm K}$ acts transitively on $M_{\rm int }$ \cite{Duff:1985jd}. A consistent truncation to a finite number of fields thus requires that the internal manifold can be represented as a coset space
\begin{equation}\label{eq:mintcoset}
M_{\rm int }=S^n =  \frac {{\rm SO}(n+1)}{{\rm SO}(n)}= \frac{{\rm K}}{{\rm L}}\,, 
\end{equation}
with ${\rm L}$ the isotropy subgroup of ${\rm K}$. 
The field content of the associated consistent truncation is given by the ${\rm L}$-singlets within the field content of the maximal supergravity. In particular, the scalar target space of the truncation is given by
\begin{equation}
{\cal M}_{\rm scalar} = \frac{{\rm Com}_{\rm L}({\rm E}_{d(d)})}{{\rm Com}_{\rm L}({\rm K}_{d})}\,,
\label{eq:cosetCEKintro}
\end{equation}
where ${\rm Com}_{{\rm L}}(G)$ denotes the commutant of ${\rm L}$ within $G$.
In the notation of \cite{Cassani:2019vcl}, ${\rm L}$ is the reduced structure group of the exceptional generalised geometry
and the intrinsic torsion is a constant ${\rm L}$-singlet, reflecting the consistency of the truncation.
When the action of ${\rm K}$ is not transitive, the truncation to ${\rm K}$-singlets contains an infinite number of fields and the
internal space is not of the form (\ref{eq:mintcoset}), but rather becomes a foliation of ${\rm K}/{\rm L}$ over another space $X$ with the coordinates on $X$ parametrising the family of singlets kept in the truncation \cite{Blair:2024ofc}. 

In general, the consistent truncation to ${\rm K}$-singlets around a sphere is not a subtruncation of the maximal supergravity. From the perspective of the maximal supergravity, they contain higher KK modes. However, 
the coset structure of (\ref{eq:mintcoset}) can be combined with the twist matrix of the maximally supersymmetric truncation in order to construct the generalised frame of exceptional field theory. This leads to fairly compact formulas for the resulting Kaluza-Klein mass matrices. In particular, the mass spectrum can still be computed in a convenient basis of scalar harmonics organised under ${\rm SO}(n+1)$, even though the actual vacuum breaks this group to the smaller group ${\rm K}$. This structure was exploited in \cite{Duboeuf:2022mam,Duboeuf:2023dmq} to compute the full KK spectrum around the squashed seven-sphere \cite{Awada:1982pk,Duff:1983ajq} represented as
\begin{equation}\label{eq:cosetSp2Sp1}
M_{\rm int }= S^7_{\rm squashed} =   \frac{{\rm Sp}(2)}{{\rm Sp}(1)}\,.
\end{equation}
Another well known example is the ${\cal N}=4$ truncation of type IIB that comes from viewing $S^5$ as a Sasaki-Einstein space \cite{Cassani:2010uw,Skenderis:2010vz,Gauntlett:2010vu}. The $\SU 2$ structure group in this case, corresponds to the $\SU 2$ denominator in the coset $S^5= {\SU 3}/{\SU 2}$.

In this paper, we review the relation between truncations to ${\rm K}$-singlets and exceptional field theory and exploit the structure in order to construct new G$_2$-invariant AdS$_4$ solutions of $D=11$ supergravity compactified on squashed seven-spheres. Specifically, we consider the truncation of $D=11$ supergravity to singlets under the G$_2$ subgroup of the ${\rm SO}(8)$ isometry group of the round $S^7$. Since ${\rm G}_2$ does not act transitively on $S^7$, the induced consistent truncation contains an infinite number of fields. For the description within exceptional field theory, we represent the seven sphere as a foliation of $S^6 ={{\rm G}_2}/{{\rm SU}(3)}$ over an interval ${\cal I}$
\begin{equation}\label{eq:S7squashed2}
M_{\rm int }=\Sigma_7 = {\cal I} \times \frac{{\rm G}_2}{{\rm SU}(3)}\,.
\end{equation}
The resulting truncation then takes the form of an ${\cal N}=2$ four-dimensional supergravity with all fields depending on an additional internal coordinate $w\in{\cal I}$, parametrising the infinite families of KK states. In particular, the spin-2 and spin-1 towers are described by a metric $g_{\mu\nu}(x,w)$, and two vector fields $A^a_\mu(x,w)$, $a=1, 2$, with $x$ denoting the AdS$_4$ coordinates. The scalar fields parametrise the coset space
\bea
{\cal M}_{\rm scalar}= \frac{{\rm SU}(2,1)}{{\rm U}(2)} \times \frac{{\rm SU}(1,1)}{{\rm U}(1)}
\;,
\label{eq:cosetG2intro}
\eea
while still depending on the extra coordinate $w$. We show that this truncation is in fact a rewriting of $D=5$ minimal gauged supergravity coupled to one hypermultiplet. In turn, this is the theory obtained by consistent truncation of $D=11$ supergravity to the (finitely many) ${\rm G}_2$-singlets on an internal $S^6$.

Searching for AdS$_4$ solutions within this truncation, we set $g_{\mu\nu}(x,w)=g_{\mu\nu}^{{\rm AdS}_4}(x)$, $A_\mu^a=0$, and restrict to scalar fields independent of the AdS coordinates $x$. We provide explicit uplift formulas for these fields to $D=11$ dimensions which produces the most general G$_2$-invariant AdS$_4$ ansatz in $D=11$ supergravity.\footnote{
%%%
Earlier constructions \cite{Gunaydin:1983mi,deWit:1984nz} were restricted to solutions living within the consistent truncation to ${\cal N}=8$ supergravity.
%%%
}
The field equations result in a system of second order ordinary differential equations for the $w$-dependent scalar fields. The system is singular at the endpoints of the interval ${\cal I}$, and we find that imposing regularity reduces its solutions to a finite discrete set.
Among them, we recover the known analytic solutions \cite{Duff:1983gq,Englert:1982vs,deWit:1984va,deWit:1984nz} which all 
live within the consistent truncation to ${\cal N}=8$ supergravity \cite{deWit:1982bul} and correspond to the 
four G$_2$-invariant extremal points of its scalar potential \cite{Warner:1983vz}. 
On top of these solutions, we identify three new regular numerical solutions of the system. Their uplift yields geometries of the form AdS$_4\times \Sigma_7$ where $\Sigma_7$ is a deformed seven-sphere, preserving SO(7) isometries, 
together with a non-vanishing three-form flux which preserves G$_2 \subset {\rm SO}(7)$ symmetry.
The analysis suggests that this is the complete set of G$_2$-invariant AdS$_4$ solutions of $D=11$ supergravity.
The description of these solutions within ExFT paves the way for a future analysis of their stability, mass spectra, supersymmetry, etc., which we leave for future work.

The rest of the paper is organised as follows. In section~\ref{sec:consistentK} we  revisit the maximal consistent truncations and the truncations to ${\rm K}$-singlets in the ExFT framework. In section~\ref{sec:G2}, we apply the construction to the truncation of $D=11$ supergravity to singlets under the G$_2$ subgroup of the ${\rm SO}(8)$ isometry group of the round $S^7$. We recover the four known G$_2$-invariant solutions and present three new numerical solutions. We close in section~\ref{sec:conclusion} with some concluding remarks and outlook.

\section{Consistent truncations to ${\rm K}$-singlets in ExFT}
\label{sec:consistentK}

In this section, we review the construction of maximal consistent truncations and consistent truncations 
to ${\rm K}$-singlets in the framework of ExFT.

\subsection{ExFT and maximal consistent truncations}

Exceptional field theory provides a reformulation of $D=11$ and IIB supergravity in terms of new variables that mimic the field content of the lower-dimensional maximal supergravity. As such, it offers a natural description of the consistent truncation of $D=11$ supergravity to the lower-dimensional maximal supergravity. For the purpose of this paper, and in particular the construction of AdS$_4$ solutions, we will focus on the E$_{7(7)}$ ExFT, constructed in \cite{Hohm:2013pua,Hohm:2013uia} to which we refer for details. Its bosonic field content comprises a $4\times 4$ metric $g_{\mu\nu}$, $\mu, \nu= 0, \dots, 3$, a set of 56 vector fields $A_\mu{}^M$, $M=1, \dots, 56$, and scalar fields parametrising a coset representative ${\cal V}$ of ${\rm E}_{7(7)}/{\rm SU}(8)$. The latter encodes the $D=11$ fields according to
\begin{equation} \label{V56}
	\begin{split}
		{\cal V} &\equiv
		{\rm exp}\left[A_{klmnpq}\, t_{(+4)}^{klmnpq} \right]
		{\rm exp}\left[A_{kmn}\,t_{(+2)}^{kmn}\right]
		\,V_{{\rm GL}(7)} \,,
	\end{split}
\end{equation}
where $V_{{\rm GL}(7)} \in {\rm GL}(7)\subset {\rm E}_{7(7)}$ is proportional to the internal block of the 11D vielbein, and $A_{kmn}$ and $A_{klmnpq}$ are the internal components of the $D=11$ three-form and dual six-form, respectively, with indices $k, l, m = 1, \ldots, 7$.
The $t_{(+n)}$ are the E$_{7(7)}$ generators of positive grading $+n$ in the algebra decomposition
\begin{eqnarray}
\mathfrak{e}_{7(7)} &{\longrightarrow}& 7'_{-4}  \oplus 35_{-2} \oplus \mathfrak{gl}(7)_0 \oplus 35'_{+2} \oplus  7_{+4} \,.
\end{eqnarray}
The remaining fields of $D=11$ supergravity parametrise the 56 vector fields $A_\mu{}^M$ and the external metric $g_{\mu\nu}$.
For later use, we state the relevant parts of the ExFT Lagrangian
\bea
g^{-1/2}\, {\cal L}_{\rm kin} 
 &=& \frac{1}{48}\,g^{\mu\nu}\,{\cal D}_{\mu}{\cal M}^{MN}\,{\cal D}_{\nu}{\cal M}_{MN}
 \;,
 \nonumber\\
g^{-1/2}\,  {\cal L}_{\rm pot} &=& 
  \frac{1}{48}{\cal M}^{MN}\partial_M{\cal M}^{KL}\,\partial_N{\cal M}_{KL}
 -\frac{1}{2} {\cal M}^{MN}\partial_M{\cal M}^{KL}\partial_L{\cal M}_{NK}
+\frac{1}{2}\,g^{-1}\partial_Mg\,\partial_N{\cal M}^{MN}
  \nonumber\\
 &&{}
  +\frac{1}{4}\,  {\cal M}^{MN}g^{-2}\partial_Mg\,\partial_Ng
  +\frac{1}{4}\,{\cal M}^{MN}\partial_Mg^{\mu\nu}\partial_N g_{\mu\nu}
\;,
\label{eq:LExFT}
\eea
in terms of the external metric $g_{\mu\nu}$, its determinant $g={\rm det} \, g_{\mu\nu}$, and the internal metric
${\cal M}={\cal V}{\cal V}^T$\,.  
%%%

The E$_{7(7)}$ ExFT formulation of $D=11$ supergravity allows for a natural description of the consistent truncation on the round $S^7$ 
to $D=4$ maximal gauged supergravity \cite{deWit:1982bul,deWit:1986iy}. 
In particular, the embedding into $D=11$ supergravity of the 70 scalar fields of $D=4$, ${\cal N}=8$ supergravity, parametrising an ${\rm E}_{7(7)}/{\rm SU}(8)$ coset representative $V$ is given by
\begin{equation}
{\cal V}(x,y) = \mathring{U}(y) \,V(x) \,,
\label{eq:VUV}
\end{equation}
in terms of the variables (\ref{V56}).
Here, $x$ and $y$ denotes the four-dimensional coordinates, and the coordinates of the seven-sphere, respectively. The 
${\rm SL}(8)\subset {\rm E}_{7(7)}$ valued twist matrix $\mathring{U}(y)$ which encodes the embedding (\ref{eq:VUV}) is  explicitly 
given by \cite{Lee:2014mla,Hohm:2014qga}
\begin{equation}
	\mathring{U}_{\underline{m}}{}^{a}({\cal Y}) =
	\begin{pmatrix}
	\mathring\omega^{3/4}\left({\cal Y}^a 
-6\,\zeta^n \partial_n {\cal Y}^a \right) \\
	 \mathring\omega^{-1/4}\,\partial_m {\cal Y}^a
	\end{pmatrix}  \in {\rm SL}(8)
	\,,\qquad
	\underline{m}=\{0,m\}\,, \quad a = \{1, \ldots, 8\} \,,
	\label{eq:Uround_alpha}
\end{equation}
in terms of the geometric data of the round $S^7$, specifically the fundamental sphere harmonics ${\cal Y}^a{\cal Y}^a=1$, the vector field $\zeta^k$ satisfying $\mathring\nabla_k\zeta^k =1$, and $\mathring{\omega}= \sqrt{ {\rm det} \,\mathring{g}_{mn}^{S^7}}$\,.
After embedding this matrix $\mathring{U}$ into the 56-dimensional fundamental representation of ${\rm E}_{7(7)}$, its
algebra valued currents
\begin{equation}
\Gamma_{\underline{M}\underline{N}}{}^{\underline{K}}  =
\rho^{-1}\,(U^{-1})_{\underline{M}}{}^P\,(U^{-1})_{\underline{N}}{}^L\,\partial_{P} U_L{}^{\underline{K}} \,,\qquad
\rho=\mathring\omega^{-1/2}
\,,
\label{eq:Gamma}
\end{equation}
define the constant intrinsic torsion
\begin{equation}
X_{\underline{MN}}{}^{\underline{K}} = -7\,\left[\Gamma_{\underline{MN}}{}^{\underline{K}}\right]_{\bf 912} \,,
\label{eq:XGamma}
\end{equation}
after projection onto the irreducible ${\bf 912}$ representation.

As an interesting consequence of the ExFT formulation, the embedding (\ref{eq:VUV}) of $D=4$ maximal supergravity can be extended to the higher Kaluza-Klein scalar modes around the round $S^7$ as
\begin{equation}
{\cal V}(x,y) = \mathring{U}(y) \,V(x) \,{\rm exp}\left[ \cP_{I} \, j^{I,\Sigma}(x) \,  \cY_\Sigma(y) \right]
,
\label{eq:scalarFluc}
\end{equation}
where the index $\Sigma$ labels the
scalar harmonics $\cY_\Sigma(y)$ on $S^7$, and the 70 non-compact generators of ${\rm E}_{7(7)}$ are denoted by ${\cal P}_I$.
In terms of ${\rm SO}(8)$ representations, these correspond to
\begin{equation}
\Sigma: \,\bigoplus_n [n,0,0,0]\;,\qquad
I:\, [2,0,0,0] \oplus [0,0,2,0]
\label{eq:repsSO8}
\end{equation}
The scalar fluctuations (including the Goldstone modes) thus fill the tensor product of these representations.
In particular, the fluctuation ansatz (\ref{eq:scalarFluc}) allows to straight-forwardly derive 
universal and compact mass formulas for the full KK spectrum of scalar fluctuations and higher point couplings~\cite{Malek:2019eaz,Malek:2020yue,Cesaro:2020soq,Duboeuf:2023cth}.

%%%%%%%%%%%%%%%%%%%%%%%%%%%%%%%%%%%%%%%%%%%%

\subsection{Consistent truncations to ${\rm K}$-singlets} 
\label{sec:Ksinglets}

We have reviewed in the above subsection, how the maximal consistent truncations on round spheres
are naturally described within the ExFT formulation of higher-dimensional supergravity. Their consistency is based on 
the underlying exceptional geometry together with the constant intrinsic torsion~(\ref{eq:XGamma}).
As discussed in the introduction, there is a different class of consistent truncations to singlets under 
some subgroup ${\rm K}$ of the isometry group of the round sphere whose consistency is ensured by
a simple symmetry argument~\cite{Duff:1985jd}. Let us briefly review how these truncations fit into the above framework.

Within the scalar sector, a truncation to the ${\rm K}$-singlets in the spectrum can be described by restricting the 
fluctuation ansatz \eqref{eq:scalarFluc} according to
\begin{equation} 
{\cal V}(x,y) 
 = \mathring{U}(y)\;\Phi (x,y)
\,,
\label{eq:TruncationSinglets}
\end{equation}
with $\Phi (x,y)$ given by
\begin{equation}
	\Phi (x,y)= {\rm exp}\Big[   \sum_{\mbox{\scriptsize K-singlets}}  \phi^\sigma(x) \, s_\sigma^{I,\Sigma}\, \cP_{I} \cY_\Sigma(y) \Big]\,.
\end{equation}
The index $\sigma$ here labels the ${\rm K}$-singlets found in the tensor product of the 
SO(8) representations (\ref{eq:repsSO8}), thereby defining the constant tensor $s_\sigma^{I,\Sigma}$.
The statement that $ \phi^\sigma(x) \, s_\sigma^{I,\Sigma}$ be $\rK$-singlet fields corresponds to 
\begin{equation}\label{eq:Kequiv}
	 \phi^\sigma(x) \, s_\sigma^{I,\Sigma}\, (k^{-1})_I {}^J =  \phi^\sigma(x) \, s_\sigma^{J,\Lambda}\, k_{\Lambda}{}^\Sigma, 
\end{equation}
where $k \in \rK$. Upon contracting with the harmonics $\cY _\Sigma$, \eqref{eq:Kequiv} corresponds to the following equivariance condition for $\Phi$
\begin{equation}\label{eq:equivscal}
	k \, \Phi(x,y) \, k^{-1} =  \Phi (x, k\cdot y), 
\end{equation}
where ``$k\,\cdot$'' is the action of $\rK$ on $M_{\rm int}$. 
I.e., the relevant $\rK$-invariant fields correspond to $\rK$-equivariant functions on the internal space $M_{\rm int}$.

In case the group ${\rm K}$ acts transitively on the internal manifold $S^7$, the number of singlets is finite,
and the truncation (\ref{eq:TruncationSinglets}) can be explicitly represented in terms of a coset representative
$S(y)$ of the internal space (\ref{eq:mintcoset})
\begin{equation}
M_{\rm int } =   \frac{{\rm K}}{{\rm L}}\,, 
\end{equation}
where ${\rm L}$ is the isotropy subgroup of ${\rm K}$. This is best understood through the language of $G$-structures as one typically does in generalised geometry \cite{Cassani:2019vcl}.

\subsection{Consistent truncation via the $\rL$-structure}

The truncation to $\rK$-singlets, as described in the previous section is consistent by the usual argument of singlets not sourcing non-singlets in the equations of motion \cite{Duff:1985jd}. Despite its simplicity, this picture has some drawbacks. It is non-trivial to identify the field content of the truncated theory from the $\rK$-singlets point of view. This applies in particular to the scalar target space and the gauging of the lower-dimensional supergravity. 
On the other hand, both the gauging and the scalar coset space can be computed systematically using the $G$-structure and intrinsic torsion data of generalised geometry \cite{Cassani:2019vcl}. In this section, we will review how the truncation to $\rK$-singlets corresponds to an ``ordinary'' truncation arising from an appropriate $\rL$-structure in generalised geometry. Let us first consider the case of finite truncations, i.e.\ the case, when the $\rK$-action on the internal space $M_{\rm int}=  \rK/\rL$ is transitive.

 Crucially, we always work on internal spaces which also admit maximal truncations, associated to a generalised frame denoted by 
 \begin{equation}
	 	\mathring {\cal U} _{\ov M}{}^M= \rho ^{-1}\, (\mathring U ^{-1})_{\ov M}{}^M\,.
        \label{eq:UU}
 \end{equation}
The truncation consists of all KK modes in the maximal theory which are invariant under $\rK$. 
Since (\ref{eq:UU}) defines a generalised parallelisation, the generalised tangent bundle $E$, its dual $E^*$, and all their tensor powers are trivial. So we can view the space of sections $\Gamma (E)$ as 
\begin{equation}
	\Gamma (E) \cong C^\infty (M)\times R_1\,, 
\end{equation}
where $R_1$ is the relevant $\En d$ representation. Analogous statements hold for the various tensor powers by replacing $R_1$ with other $\En d$ representations. In this sense, sections can be defined by simply prescribing a set of well-defined functions, valued in the relevant $\En d$ representation.

The $\rK$-invariant modes in the truncation ansatz correspond to $\rK$-equivariant functions multiplying $\mathring {\cal U}$, as seen explicitly for the scalar sector in \eqref{eq:TruncationSinglets}, \eqref{eq:equivscal}. The remaining fields obey analogous equivariance conditions.
As anticipated above, it is natural to introduce a local coset representative $S: \frac \rK \rL \rightarrow \rK$, which obeys the fundamental property 
\begin{equation}\label{eq:fundpropS}
	S(k \cdot y)= k\, S(y)\, \ell (k,y)\,,
\end{equation}
where $k\in \rK$ and $\ell(k,y)\in \rL$. The local frame 
\begin{equation}
	{\cal U}_{\ov M}{}^M= (S^{-1})_{\ov M}{}^{\ov N}\, \mathring {\cal U}_{\ov N}{}^M\, , 
\end{equation}
then defines an $\rL$-structure on $M_{\rm int}$ and allows to read off the change of basis between the global frame, and the local $\rL$-frame. Namely, if $V$ are the components of a section in the $\rK$-basis, the corresponding global function $\mathring V$ in the parallelisation basis is 
\begin{equation}\label {eq:globtoH}
	\mathring V=S\cdot V\,,
\end{equation}
where ``$\cdot$'' acts in the relevant $\En d $ representation of $V$. We can then show that $\mathring V$ is $\rK$-equivariant if and only if $V$ is a constant $\rL$-singlet. 

First, notice that $S$ is constructed by picking a fixed point $\widehat n \in\frac \rK\rL $, and building a $\rK$-valued local function $S:\frac \rK\rL\rightarrow \rK$ such that 
\begin{equation}\label{eq:Sdef}
	S(y)\cdot \widehat n = y \, ,
\end{equation}
where $y\in \frac \rK\rL$.\footnote{Note that this definition automatically implies \eqref{eq:fundpropS}, where $\rL$ is the subgroup of $\rK$ that fixes $\widehat n$.} Taking $V$ to be a constant  $\rL$-singlet,\footnote{In this context, constant means independent of $M_{\rm int}$. $V$ can still depend on the external space.} a quick calculation shows that $\mathring V$ in \eqref{eq:globtoH} is $\rK$-equivariant. 
For the converse, assume $\mathring V$ to be $\rK$-equivariant, i.e.
\begin{equation}
	k\cdot \mathring V (y)= \mathring V (k \cdot y)\,.
\end{equation}
Then by definition $V= S^{-1}\cdot \mathring V$, and it follows that $V$ is constant,
\begin{equation}
	\begin{split}
		V(y) &= S(y)^{-1}\cdot \mathring V(y)\\
		&=\mathring V\left(S(y)^{-1}\cdot y\right)= \mathring V(\widehat n)\,.
	\end{split}
\end{equation}
It is then straightforward to show that $V$ is also an $\rL$-singlet. Let $\ell\in \rL$, then 
\begin{equation}
	\ell \cdot \mathring V(\widehat n)= \mathring V (\ell \cdot \widehat n)= \mathring V (\widehat n). 
\end{equation}
We have thus shown the equivalence
\begin{equation}\label{eq:GequivHsing}
	\left\{\text{$\rK$ equivariant functions multiplying $\mathring {\cal U}$}\right\}\;\;
    \longleftrightarrow\;\;\left\{\text{Constant $\rL$-singlets multiplying ${\cal U}$}\right\}.
\end{equation}
Evaluating \eqref{eq:globtoH} on the scalar sector straightforwardly reproduces the truncation ansatz of \cite{Duboeuf:2023dmq}
 \begin{equation} 
{\cal V}(x,y) 
 = \mathring{U}(y)\;S(y)\,W(x)\,S^{-1}(y)
\;,
\label{eq:TruncationCoset}
\end{equation}
where 
 \begin{equation}
W(x)\in \frac{{\rm Com}_{\rm L}({\rm E}_{d(d)})}{{\rm Com}_{\rm L}({\rm K}_d)}\,,
\label{eq:cosetCEK}
\end{equation}
in agreement with \eqref{eq:cosetCEKintro}. Analogous expressions hold for the ans\"atze of all remaining ExFT fields. 

As discussed, the consistent truncation to finitely many ${\rm K}$-singlets in the spectrum of a round sphere $S^n={\SO{n+1}}/{\SO{n}}$, requires an alternative representation of the sphere as a coset space ${{\rm K}}/{{\rm L}}$~\cite{Duff:1985jd}. Such representations exist for a number of spheres
\cite{nlab:coset_space_structure_on_n-spheres_--_table} and Table~\ref{tab:spheresCosets} lists the examples relevant for supergravity. In the general framework of \cite{Cassani:2019vcl}, ${\rm L}$ is the reduced structure group of the exceptional generalised geometry. In all cases, the scalar coset space is given by (\ref{eq:cosetCEK}), and likewise the remaining supersymmetry of the truncation is given by the number of ${\rm L}$-singlets among the gravitini of the maximal theory. The embedding of the scalar target space into the ExFT formulation of the higher-dimensional supergravity is given by (\ref{eq:TruncationCoset}). For ${\USp{4}}/{\SU{2}}$ and $({\USp{4} \times \SU{2}})/({\SU{2}\times\SU{2}})$ this was exploited  in \cite{Duboeuf:2022mam,Duboeuf:2023dmq} to compute the full KK spectrum around the squashed seven-sphere \cite{Awada:1982pk,Duff:1983ajq}. The ${\SU{4}}/{\SU{3}}$ example has been discussed in detail in \cite{Blair:2024ofc}.

\begin{table}[bt]
	\begin{center}
		\begin{tabular}{|c|c|c|c|c|}
			\hline
			AdS $\times$ sphere & coset $\frac{{\rm K}}{{\rm L}}$ & SUSY & scalar target space & truncation \\ \hline
			AdS$_4 \times S^7$ & $\frac{\USp{4}}{\SU{2}}$ & ${\cal N}=4$ & $\frac{\SO{6,3}}{\SO{6} \times \SO{3}} \times \frac{\SL{2}}{\SO{2}}$ & \cite{Cassani:2011fu} \\[0.6em]
			AdS$_4 \times S^7$ & $\frac{\SU{4}}{\SU{3}}$ & ${\cal N}=2$ & $\frac{\SL{2}}{\SO{2}} \times \frac{\SU{2,1}}{{\rm U}(2)}$ & \cite{Gauntlett:2009zw} \\[0.6em]
			AdS$_4 \times S^7$ & $\frac{\USp{4} \times \SU{2}}{\SU{2}\times\SU{2}}$ & ${\cal N}=1$ & $\frac{\SL{2}}{\SO{2}} \times \frac{\SL{2}}{\SO{2}}$ & \cite{Cassani:2011fu,Ahn:1999dq} \\[0.6em]
			AdS$_4 \times S^7$ & $\frac{\SO{7}}{{\rm G}_2}$ & ${\cal N}=1$ & $\frac{\SL{2}}{\SO{2}}$ & \cite{Gauntlett:2009zw}  \\[0.6em]
			AdS$_4 \times S^6$ & $\frac{{\rm G}_{2}}{\SU{3}}$ & ${\cal N}=2$ & $\frac{\SL{2}}{\SO{2}} \times \frac{\SU{2,1}}{{\rm U}(2)}$ & \cite{Cassani:2009ck,Kashani-Poor:2007nby}\\[0.6em]
			AdS$_5 \times S^5$ & $\frac{\SU{3}}{\SU{2}}$ & ${\cal N}=2$ & $\frac{\SO{5,2}}{\SO{5} \times \SO{2}} \times \mathbb{R}^+$ & \cite{Cassani:2010uw,Skenderis:2010vz,Gauntlett:2010vu} \\[0.6em]
			AdS$_7 \times S^3$ & $\SU{2}$ & ${\cal N}=1$ & $\frac{\SO{3,3}}{\SO{3} \times \SO{3}} \times \mathbb{R}^+$ & \cite{Lee:2014mla} \\
			\hline
		\end{tabular}
	\end{center}
	\caption{Spheres as coset spaces and the corresponding consistent truncations.}
	\label{tab:spheresCosets}
\end{table}

Finally, let us consider a non-transitively acting $\rK$. In this case $M_{\rm int}$ is foliated into $\rK$-orbits, and a transverse space $T$\cite{Blair:2024ofc}. We denote coordinates on the orbits by $y$, and the transverse ones by $w$. It is straightforward to see that an argument identical to the transitive case still holds. However, the $\rL$-singlets in \eqref{eq:TruncationCoset} will no longer be constant, but $w$-dependent, namely 
\begin{equation}
	\begin{split}
		W (x)&\longrightarrow  W(x,w)\, . 
	\end{split}
    \label{eq:Wxqw}
\end{equation}

\subsection{Intrinsic torsion}
The gauging associated to the lower-dimensional truncated theory is encoded in the embedding tensor, which contains all the information about couplings between scalars and vectors. For truncations that arise from an $\rL$-structure, the embedding tensor is contained in the intrinsic torsion, as discussed in \cite{Cassani:2019vcl}. The above construction ensures that the intrinsic torsion of the $\rL$-structure is a constant $\rL$-singlet as we shall now sketch.
Let us begin with the case of finite trunctions. We will ignore Trombone contributions, thus restricting to $\En d$-valued generalised connections, as opposed to $\En d\times \mathbb R ^+$. It is useful to start with some general remarks.

A generalised connection $A_{M\ov N}{}^{\ov P}$ is $\rL$-compatible whenever the $A_M$ are valued in the Lie algebra $\mathfrak l$ of $\rL$. Note that barred indices are flattened with $U$, not with $\mathring U$. A covariant derivative $D_M$ is defined by 
\begin{equation}
	D_M= \del _M + (A_M\,\cdot )\,,
\end{equation} 
where $(A_M\,\cdot )$ acts in the appropriate $\mathfrak l $ representation. One can compute the torsion $\tau (D) $ of $D$\cite{Coimbra:2011ky}\footnote{Note that again we are ignoring Trombone contributions.}
\begin{equation}\label{eq:TorA}
	\tau (D)= \rho ^{-1} \, \PR{R_3} _{\ov M}{}^{\ov \alpha}{}_{\ov\beta} {}^{\ov N} \left(A_{\ov N}{}^{\ov \beta}- \Gamma_{\ov N}{}^{\ov \beta }\right)\,, 
\end{equation}
where $R_3$ is the $\En d $ representation of the embedding tensor. The greek indices in \eqref{eq:TorA} span the adjoint of $\En d$. We denote by $\Gamma$ the standard ``current'' (\ref{eq:Gamma}) associated to $U$.
One can further expand $\Gamma$ in terms of $\mathring U$ and $S $
\begin{equation}\label{eq:GammaU}
	\Gamma_{\ov {M N}}{}^{\ov P}= (S^{-1})_{\ov M}{}^{\ov R}\, (S^{-1})_{\ov N}{}^{\ov S}\, \mathring \Gamma _{\ov {RS}}{}^{\ov T}\, S_{\ov T}{}^{\ov P}+(S^{-1})_{\ov M}{}^{\ov Q}\, (S^{-1})_{\ov N}{}^{\ov S}\,\mathring  \del _{\ov Q}\, S_{\ov S}{}^{\ov P}\, , 
\end{equation}
where $\mathring \del _{\ov Q} = (\mathring U ^{-1})_{\ov Q}{}^M\, \del _M$, and $\mathring \Gamma $ is the $\mathring U$ current. 

In order to compute the intrinsic torsion, it is convenient to pick an origin in the affine space of generalised connections. One can choose 
\begin{equation}
	S^{-1}\del S\big |_{\mathfrak l}\, ,
\end{equation}
where the $\mathfrak l$ projection is taken in order to get an $\rL$-connection. Let us denote the intrinsic torsion by $\tau _{\rm int}$. We then deduce that the $R_3$ component of $\tau _{\rm int}$ is 
\begin{equation}\label{eq:Tint}
	\tau_{\rm int} =\rho ^{-1}\, \PR {R_3}\left ( -\mathring \Gamma- S^{-1}\del S\big |_{\mathfrak k\ominus \mathfrak l}\right)\, , 
\end{equation}
where all indices are suppressed. Let us make a couple of observations about \eqref{eq:Tint}: 
\begin{enumerate}
	\item The current $\mathring \Gamma$ is contracted with $S$ as in \eqref{eq:GammaU}.
	\item The derivative in $S^{-1}\del S\big |_{\mathfrak k\ominus \mathfrak l}$ is really $(S^{-1})_{\ov M}{}^{\ov N}\, \mathring \del _{\ov N}$. 
\end{enumerate}
The $\mathring \Gamma $ term in \eqref{eq:Tint} gives the constant embedding tensor $\mathring X $ of the maximal truncation associated to the global frame $\mathring {\cal U}$. It is crucial to note that $S\in \rK$ belongs to some subgroup of the gauging, thus $\mathring X $ is also a $\rK$-singlet. Hence, the $S$ dressing leaves $\mathring X$ invariant.  We can now focus on the $S^{-1}\del S$ term.

\subsubsection{Finite case}
We will specialise to the situation of interest, i.e., when the internal space is $S^n$, and $\mathring U$ is its usual parallelisation of \cite{Lee:2014mla,Hohm:2014qga}. $S$ should be viewed as a local function on the sphere. Furthermore, derivatives obey the section condition. Hence, we can write the action of $\mathring \del$ as 
\begin{equation}
	\rho ^{-1} \, \mathring \del_{\ov M} = K_{\ov M}{}^i \, \del _i\, ,
\end{equation}
where $K_{\ov M}{}^i$ are the $\SO {n+1}$ Killing vectors of the round $n$-sphere. From now on, let us introduce $\SO {n+1}$ fundamental indices $\ov a = 1,\ldots , n+1$, and denote the non-trivial Killing vectors by $K_{\ov {ab}}$. It is also convenient to introduce $\SO {n+1}$ generators 
\begin{equation}
	(T_{\ov{ab}})_{\ov c}{}^{\ov d}	= 2 \, \delta_{\ov c [\ov a}\,\delta _{\ov b ]}{}^{\ov d}\,. 
\end{equation}
The action of $K_{\ov {ab}}$ on a scalar function $f$ reduces to 
\begin{equation}\label{eq:Konf}
	K_{\ov {ab}}{}^i\, \del _i \, f(y)= \frac d {dt}\bigg |_{t=0} f\left ( e ^{tT_{\ov {ab}}} y\right )\,.
\end{equation}
More specifically, in order to match \eqref{eq:GammaU}, $\mathring \del $ must be contracted with $S^{-1}$. Hence, \eqref{eq:Konf} becomes 
\begin{equation}\label{eq:SKonf}
	\begin{split}
		(S^{-1})_{\ov {ab}}{}^{\ov {cd}}\,K_{\ov {cd}}{}^i\, \del _i \, f(y)&= \frac d {dt}\bigg |_{t=0} f\left ( S(y)\,e ^{tT_{\ov {ab}}}\, S^{-1}(y) y\right )\,,
	\end{split}
\end{equation}
where we used that $T_{\ov {ab}}$ are $\rK$-singlets. Applying \eqref{eq:SKonf} to our situation, leaves us with the following expression
\begin{equation}\label{eq:FiniteSdS}
	S^{-1}(y)\, \frac d {dt }\bigg| _{t=0} \, S\left(S(y)\, e ^{t T_{\ov {ab}}}S^{-1}(y)\, y \right)\,.
\end{equation}
By definition the coset representative satisfies $S^{-1}(y)y= \widehat n $. Hence, \eqref{eq:FiniteSdS} vanishes whenever $T_{\ov {ab}}$ is an element of the Lie subalgebra $\mathfrak {so }(n)$. Therefore, the only non-trivial contribution comes when $T_{\ov {ab}}\in \mathfrak{so }(n+1)\ominus \mathfrak {so }(n)= \mathfrak k \ominus \mathfrak l$.\footnote{This equality holds up to $\mathfrak{so }(n)$ shifts.} One can then take $T_{\ov {ab}}$ to lie in $\mathfrak k \ominus \mathfrak l$, without loss of generality. Let us label generators of $\mathfrak k \ominus \mathfrak l$ by indices $\ov I,\, \ov J,\,\ldots$. Crucially, $T_{\ov I}\in \mathfrak k$, so we can use the fundamental property of $S$ to pull out the exponential 
\begin{equation}
	S\left(S(y)\, e ^{t T_{\ov I}}S^{-1}(y)\, y \right)= S(y)\, e ^{t T_{\ov I}}\, e^{\chi \left(t,y\right )}\,,
\end{equation} 
where $\chi\left(t,y\right)$ is a curve in $\mathfrak l$ with $\chi\left (0,y\right )=0$. It is then straightforward to see that \eqref{eq:FiniteSdS} reduces to 
\begin{equation}\label{eq:Tintconst}
	\begin{split}
		S^{-1}(y)\, \frac d {dt }\bigg| _{t=0} \, S\left(S(y)\, e ^{t T_{\ov {ab}}}S^{-1}(y)\, y \right) &= \frac d {dt }\bigg |_{t=0} e ^{t T_{\ov I}}\, e^{\chi \left(t,y\right )}\\[0.4em]
		&=T_{\ov I} + \dot \chi \left (0,y\right )\, , 
	\end{split}
\end{equation}
where $\dot \chi $ is the $t$ derivative of $\chi$. To conclude, substituting \eqref{eq:Tintconst} into \eqref{eq:Tint}, shows that the remaining component of $\tau _{\rm int }$ corresponds to the $(\mathfrak k \ominus \mathfrak l)\otimes (\mathfrak k \ominus \mathfrak l )$ block of the Cartan-Killing form $\kappa$ of $\SO {n+1}$, appropriately embedded in $\En d$, and projected onto $R_3$. Crucially, $\kappa _{\ov {IJ}}$ is a constant $\rL$-singlet.

\subsubsection{Infinite case}
The infinite case is more involved. In this paper, we will restrict to the case where the transverse space $T$ is one-dimensional, such that the fields of the truncation (\ref{eq:Wxqw}) depend on one transverse coordinate $w$ only. Specifically, we view the internal space $M_{\rm int}=S^n$ as a fibration of $S^{n-1}$ over the interval ${ \cal I}=[-1,1]$.  We take $S^{n-1}\cong  \rK/ \rL$, where $\rK$ acts transitively on $S^{n-1}$. More concretely, we write the $S^{n}$ embedding coordinates as 
\begin{equation}
	{\cal Y}= (\sqrt {1-w ^2}\,y, w )\,,
\end{equation}
where $y=(y^1,\ldots , y^{n})$ are embedding coordinates of $S^{n-1}$, and $\omega $ parametrises ${\cal I}$. The coset representative $S$ is then only a function of $y$, while $\widehat n$ is replaced with a copy of ${\cal I}$. More specifically, we introduce 
\begin{equation}\label{eq:infn}
	\widehat n(w)= (\sqrt {1-w ^2}, 0, \ldots , 0, w)\,,
\end{equation}
if we then view $S$ and $\widehat n$ as (local) functions on $S^n$, the defining property \eqref{eq:Sdef} becomes\footnote{We slightly abuse notation here, one has to keep in mind that $S$ only depends on the $S^{n-1}$ coordinates. Similarly $\widehat n$ is a function of $w$ only. } 
\begin{equation}\label{eq:modifSdef}
	S({\cal Y})\, \widehat n({\cal Y}) = {\cal Y}\, .
\end{equation}
Crucially, the isotropy group $\rL$ does not stabilise a single point anymore, but a copy of ${\cal I}$. Namely, it is defined by 
\begin{equation}
	\ell \, \widehat n(\cY)=\widehat n (\cY),\text{ with } \widehat n (\cY)=(\sqrt {1-w^2}, 0, \ldots , 0, w)\,.
\end{equation}
Since $\rK$ acts on $S^{n-1}$, the modification \eqref{eq:modifSdef} still implies the universal property \eqref{eq:fundpropS}. Much of the finite case analysis still applies. In particular, the non-trivial term we should compute is still \eqref{eq:FiniteSdS}. However, there are two significant differences
\begin{enumerate}
	\item $S^{-1}(y)\, y= \widehat n$ is not a constant: it corresponds to $\widehat n (w)$ of \eqref{eq:infn}. 
	\item In the infinite case, \eqref{eq:FiniteSdS} is non-trivial \textit{also} for some generators outside of $\mathfrak k \ominus \mathfrak l$. For example, in our case, the $T$'s transforming in the vector representation of $\SO n$ give non-vanishing contributions. \label{pt:point2inf}
\end{enumerate}
Because of point \ref{pt:point2inf} above, we cannot use the fundamental property of $S$ to pull out $S\, e^{tT}\, S^{-1}$, as we did for the finite case. We instead write:
\begin{equation}\label{eq:pulloutinf}
	S\left(S(z)\, e ^{t T_{\ov ab}}S^{-1}(z)\, z \right)= S(z)\, S(e^{tT_{\ov {ab}}}\, \widehat  n (w))\, \ell (t, z)\, . 
\end{equation}
Note that here we denote the $n$ local coordinates on $S^n$ by $z$, which in turn correspond to $w$ along with the $S^{n-1}$ local coordinates. From now on, let us assume that $S\left(\widehat n (w)\right )=1$.\footnote{This can be done without loss of generality.} One can easily see that $\ell (0,z)=1$. Thus, for sufficiently small $t$, we can again assume 
\begin{equation}
	h(t,z)= e ^{\chi (t,z)},\text{ with $\chi$ some curve in }\mathfrak l\,. 
\end{equation}
We now show that in the infinite case, the $\mathfrak k \ominus \mathfrak l$ projection of $S^{-1}\del S$ \textit{only} depends on $w$, and is an $\rL$-singlet.

The $w$-dependence follows from \eqref{eq:pulloutinf}. Namely
\begin{equation}\label{eq:SdSinf}
	\begin{split}
		S^{-1}(z)\, \frac d {dt }\bigg| _{t=0} \, S\left(S(z)\, e ^{t T_{\ov {ab}}}S^{-1}(z)\, z \right) &= \frac d {dt }\bigg |_{t=0} S(e^{tT_{\ov {ab}}}\, \widehat  n (w))\,e ^{\chi (t,z)}\\[0.4em]
		&= \frac d {dt }\bigg |_{t=0} S(e^{tT_{\ov {ab}}}\, \widehat  n (w))+ \dot \chi \left (0,z\right )\, , 
	\end{split}
\end{equation}
it is clear that the $\mathfrak k \ominus \mathfrak l$ component of \eqref{eq:SdSinf}, can only depend on $w$. We will now show that \eqref{eq:FiniteSdS} is an $\rL$-singlet. 

Let $\Lambda \in \rL$, one can act directly on \eqref{eq:SdSinf}
\begin{equation}\label{eq:TintinfLambda}
	\begin{split}
		\Lambda \, \frac d {dt }\bigg |_{t=0} S\left (\Lambda ^{-1}\, e^{tT_{\ov {ab}}}\, \widehat  n (w)\right)\,e ^{\chi (t,z)}\, \Lambda ^{-1}&=	\frac d {dt }\bigg |_{t=0}S(e^{tT_{\ov {ab}}}\, \widehat  n (w))\, \lambda (t,w)\, e ^{\chi (t,z)}\Lambda ^{-1}\,, 
	\end{split}
\end{equation}
where we again use the fundamental property of $S$. The extra factor $\lambda (t,w)\in \rL$ ``compensates'' for $\Lambda^{-1} $ being pulled out of $S$ on the left-hand side. Let us now consider $t=0$, 
\begin{equation}
	\begin{split}
		1=S(\widehat n (w )) &= S(\Lambda ^{-1}\, \widehat n (w )) \\
		&= \Lambda ^{-1}\, S(\widehat n(w ))\,\lambda (0,w) \\
		&= \Lambda ^{-1}\,\lambda (0,w)\,, 
	\end{split}
\end{equation}
so that $\lambda (0,w)= \Lambda$. Hence, again for small $t$, we can take 
\begin{equation}\label{eq:Tintsinginf}
	\lambda (t,w)= \Lambda \, e ^{\xi(t,w)}\,, \text { with $\xi$ a curve in $\mathfrak l$ such that $\xi (0,w)=0$.}
\end{equation}
Substituting \eqref{eq:Tintsinginf} into \eqref{eq:TintinfLambda} gives 
\begin{equation}
	\Lambda \, \frac d {dt }\bigg |_{t=0} S\left (\Lambda ^{-1}\, e^{tT_{\ov {ab}}}\, \widehat  n (w)\right)\,e ^{\chi (t,z)}\, \Lambda ^{-1}= \frac d {dt }\bigg |_{t=0} S(e^{tT_{\ov {ab}}}\, \widehat  n (w))+ \text{ some $\mathfrak l$ contribution, }
\end{equation}
Thus we conclude that the $\mathfrak k \ominus \mathfrak l$ projection is an $\rL$-singlet. 

To conclude, in the infinite case, the intrinsic torsion is still an $\rL$-singlet. However, unlike the finite case, it is not constant. Instead, it depends on the transverse coordinate, $w$. It would be interesting to explicitly evaluate $\tau _{\rm int}$ for the infinite case.
    
\section{${\rm G}_2$-invariant solutions of $D=11$ supergravity}
\label{sec:G2}

In this section, we will use the ExFT structures in order to revisit and construct 
new ${\rm G}_2$-invariant AdS$_4 \times \Sigma_7$ solutions of $D=11$ supergravity.
Solutions of this type have been constructed in the past directly in $D=11$ dimensions
\cite{Duff:1983gq,Englert:1982vs,deWit:1984va}, and most systematically in \cite{deWit:1984nz}. 
However, all previous constructions have been restricted to solutions that live within the consistent 
truncation to ${\cal N}=8$ supergravity \cite{deWit:1982bul}. In terms of the $D=4$ theory, they correspond 
to the G$_2$-invariant extremal points of the scalar potential \cite{Warner:1983vz}. 

Instead, here we will allow for a deformation of the round $S^7$ by the most general combination of the infinitely many
${\rm G}_2$-invariant scalar modes in the KK spectrum. In the ExFT framework, this corresponds to analysing a consistent truncation
to infinitely many fields \cite{Blair:2024ofc}. Since we are interested in AdS$_4$ solutions, we focus on the scalar sector of the
four-dimensional theory.

\subsection{Consistent truncation to G$_2$-singlets}

We start by representing the seven-sphere $S^7$ as a foliation of $S^6$ over an interval ${\cal I}$. 
Specifically, we use its embedding coordinates ${\cal Y}^I$ inside $\mathbb{R}^8$, satisfying ${\cal Y}^I{\cal Y}^I=1$,
and represent them as
\begin{equation}
{\cal Y}^i = \sqrt{1-w^2}\,y^i\;,\quad {\cal Y}^8 = w\in [-1,1]\;,\quad i=1, \dots, 7
\;,
\label{eq:Yembedd}
\end{equation}
with the embedding coordinates $y^i$ of the round $S^6$ satisfying $y^i y^i=1$.
We will also introduce the angle coordinate $\theta\in[0,\pi]$ by
\begin{equation}
w=-{\rm cos}\,\theta
\;.
\label{wt}
\end{equation}
According to the general discussion, within ExFT the consistent truncation to G$_2$-singlets in the scalar sector
is described by a parametrisation of the generalised vielbein as
\begin{equation} 
{\cal V}(x,y,\theta) 
 = \mathring{U}(y,\theta)\;S(y)\,W(x,\theta)\,S^{-1}(y)
\;.
\label{eq:TruncationCosetG2}
\end{equation}
Here, $S(y)$ is a coset representative for 
\begin{equation}
S^6=\frac{{\rm G}_{2}}{\SU{3}}
\;,
\label{eq:cosetS6}
\end{equation}
and the $W$ is a coset representative of the coset (\ref{eq:cosetCEKintro})
\begin{equation}
\frac{{\rm Com}_{{\rm SU}(3)}({\rm E}_{7(7)})}{{\rm Com}_{{\rm SU}(3)}({\rm SU}(8))}
= \frac{{\rm SU}(2,1)}{{\rm U}(2)} \times \frac{{\rm SU}(1,1)}{{\rm U}(1)}
\,,
\label{eq:cosetCEKG2}
\end{equation}
still depending on the additional coordinate 
$\theta\in[0,\pi]$, parametrising the infinite families of KK states. 
We denote by $\{\phi, \chi\}$ the coordinates of the second factor of (\ref{eq:cosetCEKG2}), and parametrise the 
quaternionic manifold ${{\rm SU}(2,1)}/{{\rm U}(2)}$ by coordinates
\bea
\{\phi_1,\chi^m\} = \{\phi_1,\chi_{1a},\chi_{1b},\chi_2\}\;.
\label{coordQ}
\eea
By virtue of their $\theta$-dependence, each of these six fields represents an infinite family of four-dimensional scalars, which, however, still include both physical scalars together with the Goldstone modes.
As for the remaining bosonic fields, the truncation carries two infinite families of vector fields,
parametrised by $\theta$ as $A^\alpha_\mu(\theta)$, $\alpha=1, 2$, as well as the spin-2 tower described by
$g_{\mu\nu}(\theta)$. After Higgsing (for spin-1 and spin-2 fields), 
the theory then describes the massive spin-2 tower, together with one massive spin-1 tower 
and four infinite towers of massive scalar fields.
In the fermionic sector, the four-dimensional theory after Higgsing carries two infinite towers of massive gravitino fields $\psi_\mu^u(\theta)$, $u=1, 2$,
together with two towers of massive spin $1/2$ fermions.
Indeed, this matches the counting of ${\rm G}_2$-singlets within the KK spectrum around the round sphere $S^7$ \cite{Englert:1983rn,Biran:1983iy,Sezgin:1983ik,Casher:1984ym}.

When searching for AdS$_4$ solutions, we will impose vanishing fermions, and set 
\begin{equation}
A^a_\mu(\theta)=0\;,\qquad
\partial_\theta g_{\mu\nu}(\theta) = 0
\;.
\label{gA0}
\end{equation}
The external part ${\cal L}_{\rm kin}$ of the ExFT Lagrangian (\ref{eq:LExFT}) in this truncation is computed by
evaluating the Lagrangian with the ansatz (\ref{eq:TruncationCosetG2}), leading to
\bea
 g^{-1/2}\,{\cal L}_{\rm kin} 
  \,=\,
- 2\rho^{-2}\left(
\partial_\mu\phi_1 \partial^\mu \phi_1
+e^{-4\phi_1}\,
M_{mn}\,\partial_\mu\chi^m \partial^\mu \chi^n
+ \frac34\left(
\partial_\mu\phi \partial^\mu \phi +e^{-2\phi}\,\partial_\mu\chi \partial^\mu\chi\right)
 \right)
 ,
 \label{LkinG2}
\eea
with the weight factor
\begin{equation}
\rho=({\rm sin}\,\theta)^{-3}
\;,
\label{rho}
\end{equation}
and the scalar matrix $M_{mn}$ given by
\bea
M_{mn} =
\begin{psmallmatrix}
 e^{2\phi_1}+\chi_{1b}^2 & -\chi_{1a}\chi_{1b} &-\chi_{1b} \\[.5ex]
-\chi_{1a}\chi_{1b} & e^{2\phi_1}+\chi_{1a}^2 &\chi_{1a} \\[.5ex]
-\chi_{1b} & \chi_{1a}& 1
\end{psmallmatrix}
\;.
\label{Mij}
\eea
This confirms that the scalar kinetic term (\ref{LkinG2}) is given by a four-dimensional sigma-model on the six-dimensional target space (\ref{eq:cosetCEKG2}) with all fields carrying an additional dependence on the coordinate $\theta$.

In the search for AdS$_4$ solutions, we will further restrict to scalar fields that are constant in AdS$_4$, i.e.\ reduce to functions of only the additional coordinate $\theta$. For such solutions, the kinetic Lagrangian (\ref{LkinG2}) vanishes and does not contribute to the field equations.
The relevant Lagrangian is thus obtained from the internal part ${\cal L}_{\rm pot}$ of the ExFT Lagrangian (\ref{eq:LExFT}). After some lengthy but straightforward computation this yields the truncated Lagrangian
\bea
 {\cal L}_{\rm pot} &=& 
 \rho^{-2}\,g^{-1/2}\,L_{\rm pot}
  \;,
\eea
with
\bea
L_{\rm pot} &\!=\!&
\frac{3}{2}\,e^{-3 \phi }D_\theta\phi D_\theta \phi
-\,e^{-3 \phi }\,D_\theta\phi_1 D_\theta \phi_1
-e^{-3 \phi -4\phi_1}\,
M_{mn}\,D_\theta \chi^m D_\theta \chi^n
\label{LpotG2}\\
&&{}
+3\,e^{-\phi} \,
\Big\{2 \cot ^2\theta \left(5 \chi_{1a}^2+3 \chi_{1b}^2\right)
-4\, e^{-4 \phi_1} \left(
\chi_{1a}+\cot\theta\left(\chi_{1a}^2 \chi_{1b} +2 \chi_{1a}\chi_{2} +\chi_{1b}^3\right) \right)^2
\nonumber\\
&&{}   
\quad    + e^{-2\phi_1} \Big((5+\cot\theta\,(10\chi_2 -8 \chi_{1a} \chi_{1b}))(1+2\cot\theta\, \chi_{2})
   + \cot^2\theta\,(5 \chi_{1a}^2-3 \chi_{1b}^2)(\chi_{1a}^2+ \chi_{1b}^2)\Big)
   \Big\}
\nonumber\\
&&{}   
\quad +15\, e^{-\phi+2 \phi_1} \cot ^2\theta
- 3\,e^{-3 \phi} \,(1+3   \cos (2 \theta ))\csc^2 \theta
\;,
\nonumber
\eea
with the matrix $M_{mn}$ from (\ref{Mij}) above and the `covariant' derivatives $D_\theta$ defined as
\begin{align}
D_\theta\phi =&\;
\partial_\theta \phi+2 \cot \theta
   \;,\nonumber\\
D_\theta\phi_1 =&\;
\partial_\theta \phi_1+2\,   \chi_2 -3 \cot\theta -6\, \chi \chi_{1a} \cot \theta
   \;,\nonumber\\
D_\theta\chi_{1a}  =&\;
\partial_\theta \chi_{1a} 
-3\,\chi_{1a} \cot\theta
 -\chi_{1b}\left(e^{2 \phi_1}+\chi_{1a}^2   
+\chi_{1b}^2\right)
+ 2\,\chi_{1a}\chi_2
+3\, \chi \cot\theta 
   \left(e^{2 \phi_1} -\chi_{1a}^2+ 3\,\chi_{1b}^2\right)
  ,\nonumber\\
D_\theta\chi_{1b}  =&\; \partial_\theta \chi_{1b}
-3 \chi_{1b} \cot \theta
+\chi_{1a}\left(
e^{2 \phi_1}
+ \chi_{1a}^2
+\chi_{1b}^2\right)
+2 \chi_{1b} \chi_2
-3 \chi \left(1+2 \cot\theta\, (2\,\chi_{1a} \chi_{1b}+\chi_2)\right)
  ,\nonumber\\
D_\theta\chi_2  =&\;
\partial_\theta \chi_{2} -\tfrac12\,e^{4\phi_1}-\tfrac52
-6 \chi_2 \cot\theta
   -\tfrac12\left(\chi_{1a}^2+\chi_{1b}^2\right) \left(2\, e^{2\phi_1}+\chi_{1a}^2+\chi_{1b}^2\right)
   +2 \chi_2^2 
   \nonumber\\
   &{}\;
+3 \chi\left( \cot\theta     \left(
\chi_{1b}   \left(e^{2 \phi_1}+\chi_{1a}^2+\chi_{1b}^2\right)-2\chi_{1a} \chi_2\right)-\chi_{1a}\right)
  . 
  \label{Dtheta}
\end{align}
All fields depend on the coordinate $\theta$ only, variation of (\ref{LpotG2}) thus implies a set of 
ordinary differential equations for the scalar fields.
Furthermore, the Lagrangian (\ref{LpotG2}) explicitly depends on the coordinate $\theta$
induced by the $\theta$-dependence of the generalised frame of the round sphere $\mathring{U}$
in the truncation ansatz (\ref{eq:TruncationCosetG2}).

It is  straightforward to check that the combination
\begin{equation}
V = -{L}_{\rm pot}-\frac34\,\rho^{2}\, \partial_\theta\left( \rho^{-2}\,e^{-3\phi}\,\partial_\theta\phi \right)
\;,
\label{V}
\end{equation}
is conserved on-shell, i.e.\ $\partial_\theta V=0$ as a result of the field equations implied by (\ref{LpotG2}). This charge shows up in the 
four-dimensional Einstein field equations in this truncation,
and encodes the AdS$_4$ radius $\ell_4$ as
\bea
R_{\mu\nu}-\frac12\,g_{\mu\nu}\,R = -V\,g_{\mu\nu} = \frac3{\ell_4^2}\,g_{\mu\nu}
\;.
\label{ell4}
\eea

The uplift of the consistent truncation can be computed by extracting the $D=11$ fields upon combining 
(\ref{V56}) with the ansatz (\ref{eq:TruncationCosetG2}). For instance, the $D=11$ metric is expressed in terms of the scalar fields (\ref{coordQ}) as
\bea
ds^2 &=& \Delta^{-1}\,ds_{(4)}^2
+ e^{3\phi}\,\Delta^{-1}\,d\theta^2
+e^{-\frac{\phi}{2}}\,{\rm sin}^2\theta\,\Delta^{1/2}\,ds_{S^6}^2
\;.
\label{metric11DG2}
\eea
Here, $ds_{S^6}^2$ denotes the metric of the round $S^6$, and the warp factor $\Delta$
is given by
\bea
\Delta &=& e^{\phi+4\phi_1/3} \,({\rm cos}\,\theta)^{-4/3} \left(
\left(e^{2\phi_1}+\chi_{1a}^2+\chi_{1b}^2\right)^2+\left({\rm tan}\,\theta+2\,\chi_2\right)^2\right)^{-2/3}
\;.
\label{DeltaG2}
\eea
All fields are functions of $\theta$. 
Similarly, one may extract the $D=11$ three-form.
We give explicit formulas below, c.f.\ (\ref{11DfieldsG2}), after further simplification of the system.

By construction, every solution to the equations of motion of (\ref{LpotG2}) locally describes a solution of $D=11$ supergravity.
However, searching for a solution with compact internal space given by a seven-sphere, the form of the metric (\ref{metric11DG2})
shows that we need to require all fields to remain regular at the endpoints of the interval $\theta\in[0,\pi]$.
As we will make explicit below, the field equations derived from (\ref{LpotG2}) are singular at these endpoints, such
that the proof of existence and the construction of regular solutions becomes a rather non-trivial task. In particular, we will see that only
a discrete and finite set of such solutions exists.

Before proceeding with the analysis of solutions, let us note that the 
Lagrangian (\ref{LpotG2}) can be further simplified. First, we observe that it
is invariant under the gauge transformations
\bea
\delta \chi &=& -\partial_\theta\Lambda -2\,\Lambda\,\cot\theta
\;,\nonumber\\
\delta_\Lambda\phi_1 &=& -6\, \Lambda\, \chi_{1a} \cot \theta
   \;,\nonumber\\
\delta_\Lambda\chi_{1a}  &=&
3\, \Lambda\, \cot\theta 
   \left(e^{2 \phi_1} -\chi_{1a}^2+ 3\,\chi_{1b}^2\right)
  \;,\nonumber\\
\delta_\Lambda\chi_{1b}  &=&
-3 \,\Lambda \left(1+2 \cot\theta\, (2\,\chi_{1a} \chi_{1b}+\chi_2)\right)
  \;,\nonumber\\
\delta_\Lambda \chi_2  &=&
3 \,\Lambda\left( \cot\theta     \left(
\chi_{1b}   \left(e^{2 \phi_1}+\chi_{1a}^2+\chi_{1b}^2\right)-2\chi_{1a} \chi_2\right)-\chi_{1a}\right)
  \;,
  \label{deltaLambda}
\eea
with arbitrary $\Lambda=\Lambda(\theta)$, which can be used to eliminate one of the scalar fields. 
Moreover, the Lagrangian (\ref{LpotG2}) only depends algebraically on the field $\chi$ which 
can thus be integrated out by virtue of its field equations, reducing the system to only four scalar fields.
This is a remnant of the Higgs mechanism of the full theory.

As it turns out, the system can further be drastically simplified by going to different variables which are closer to the higher-dimensional origin of the fields. We will show in the following that upon change of coordinates and fields, the Lagrangians (\ref{LkinG2}), (\ref{LpotG2}) embed into a simple five-dimensional Lagrangian upon merging the AdS$_4$ coordinates $x$ and the extra coordinate $\theta$ into a five-dimensional space-time.
In turn, this significantly simplifies the equations of motion such that they can be treated by numerical methods.
In order to illustrate this simplification of the system, we will first discuss the further subtruncation to the (still infinitely many) ${\rm SO}(7)$-singlets in the spectrum of the round sphere.

\subsection{${\rm SO}(7)$ truncation}
\label{sec:SO7}

As an illustration, let us first discuss the truncation of the system to 
the ${\rm SO}(7)$-singlets in the $S^7$ spectrum. 
This sector has been discussed in \cite{Blair:2024ofc}.
Within the above discussion, this corresponds to the further (consistent) truncation
\bea
\chi=\chi_{1a}=\chi_{1b}=0
\;,
\label{truncSO7}
\eea
such that we are left with three $\theta$-dependent scalar fields $\{\phi, \phi_1, \chi_2\}$ parametrising the target space
\bea
\frac{{\rm SU}(1,1)}{{\rm U}(1)} \times \mathbb{R}
\;,
\eea
with $\phi$ denoting the $\mathbb{R}$ coordinate.
The ExFT action (\ref{LkinG2}), (\ref{LpotG2}) in this truncation reduces to
\bea
S&=& \int d^4x\,d\theta\,\rho^{-2}\,g^{-1/2}\,
\left({L}_{\rm kin} + {L}_{\rm pot} \right),
\label{SExFT}
\eea
with 
\bea
{L}_{\rm kin}&=&
-\frac32\,\partial_\mu\phi \partial^\mu \phi 
-2\,\partial_\mu\phi_1 \partial^\mu \phi_1
-2\,e^{-4\phi_1}\,\partial_\mu\chi_2 \partial^\mu \chi_2
\;,
\label{LkinSO7}
\\[2ex]
{L}_{\rm pot} &=&
\frac32  \, e^{-3 \phi } \left(\partial_\theta \phi+2 \cot \theta \right)^2
- e^{-3 \phi } \left(\partial_\theta\phi_1+2 \chi_2-3 \cot  \theta \right)^2
\label{LpotSO7}
\\
&&{}
- e^{-3 \phi -4\, \phi_1} \Big(\partial_\theta\chi_2
   +2   \chi_2^2-\tfrac12 e^{4 \phi_1}-6\, \chi_2 \cot  \theta-\tfrac52 \Big)^2
\nonumber\\
&&{}
+
15 \,e^{-\phi -2\phi_1} (1+2 \,\chi_2\, \cot \theta)^2
+15 \,\cot^2\theta\,e^{- \phi +2 \phi_1}
-2\,e^{-3 \phi }\,(1+3   \cos (2 \theta )) \csc ^2 \theta 
   \Big)
\;.
\nonumber
\eea
Again, one finds that the combination $V$ from (\ref{V}) is conserved on-shell, 
i.e.\ as a result of the field equations implied by (\ref{LpotSO7}),
and encodes the AdS$_4$ radius $\ell_4$ according to (\ref{ell4}).\footnote{
After change of coordinates
\begin{equation}
\phi_1\rightarrow-\tfrac12\varphi\;,\quad
\chi_2\rightarrow\chi\;,\quad
\theta\rightarrow\pi-\theta_7
\;,
\end{equation}
one may further check that $V$ in the ${\rm SO}(7)$ truncation
precisely reproduces the effective potential derived in \cite{Blair:2024ofc}, 
where it is directly obtained via the embedding (\ref{eq:TruncationCosetG2}) with (\ref{eq:cosetS6}) 
replaced by ${{\rm SO}(7)}/{{\rm SO}(6)}$ and given in their equation (3.31).
}
Furthermore, one finds that the field equations obtained from (\ref{LpotSO7}) imply that
\begin{equation}
\Delta^{-3}\,\rho^2\,\partial_\theta\left(
\rho^{-2}\,e^{-3\phi}\,\Delta^2\,C_\theta
\right) = {\rm const}
\;,
\label{dCc}
\end{equation}
where $C_\theta$ is defined by 
\begin{equation}
 C_\theta  =
{\rm tan}\,\theta\,e^{3\phi}\,\Delta^{-2}\left(1-f(\theta)\right)
-e^{\frac{3\phi}{2} - 2\phi_1}\,\Delta^{-1/2}
\left({\rm tan}\,\theta+2\,\chi_2\right)
\;,
\label{C311D}
\end{equation}
with the function $f$ satisfying the differential equation
\bea
f'(\theta) &=& 
\left(6-7f(\theta)\right){\rm cot}\theta - f(\theta)\,{\rm tan}\,\theta
\;.
\label{deff}
\eea
Equation (\ref{dCc}) is inherited from the Bianchi identity in $D=11$ supergravity, with $C_\theta$
describing the $D=11$ internal 6-form as $\star_7\, C_{(6)} = C_\theta \, d\theta$\,.

The existence of two conserved charges (\ref{V}) and (\ref{dCc}) suggests that the model (\ref{SExFT})
can be further simplified exploiting the associated global symmetries. Indeed this is the case and 
leads to a compact reformulation in terms of redefined coordinates and fields that are naturally
associated with a $D=5$ uplift of the field equations as we shall show now.

\subsubsection{Redefined coordinates and fields}

Exploiting the global symmetry derived from the conserved charge (\ref{V})
reveals a redefinition of the $\theta$-coordinate, which together with a $\theta$-dependent 
$\mathbb{R}\times {\rm SL}(2)$ transformation on the scalar fields,
leads to a Lagrangian which no longer shows any explicit coordinate-dependence.
Explicitly, this is achieved, by going to the coordinate $u$ defined as
\bea
u&=& -\frac{1}{96}\, \Big( 96+ 45\, \sin (2 \theta )-9\, \sin (4 \theta )+\sin (6 \theta ) -60\, \theta \Big)
\;\;\in\; [-1,1]
\;,
\nonumber\\
&&\Longrightarrow\quad
\partial_\theta u=2\, \rho^{-2}=2\, {\rm sin}^6\theta
\;.
\label{defu}
\eea
Simultaneously, we define the scalar fields as
\bea
\{\phi, \phi_1, \chi_2\} &\longrightarrow&
\{\Phi, \Phi_1, {\cal X}_2\}
\;,
\nonumber\\[2ex]
\mbox{with}\qquad
e^{\Phi} &=& \rho^{4/3}\,e^{\phi}
\;,
\nonumber\\
e^{\Phi_1} &=& \rho^{-1}\,e^{-3\phi/4} \,\Delta^{3/4}
\;,
\nonumber\\
{\cal X}_2 &=&
\frac1{2}\,\rho^{-2}\,
\left(
e^{-3\phi/2-2\phi_1}\,\Delta^{3/2} \left( {\rm tan}\,\theta+2\,\chi_2\right)
-\left({\rm tan}\,\theta-3\,\rho^2\,u\right)
\right)\;
,
\label{5DNewFields}
\eea
with $\rho$, $\Delta$, and $u$ from (\ref{rho}), (\ref{DeltaG2}), and (\ref{defu}), respectively.
Although not manifest, one may verify that this transformation, corresponds to a
non-linear $\theta$-dependent $\mathbb{R}\times {\rm SU}(1,1)$ transformation on the fields. As a result, 
the kinetic term (\ref{LkinSO7}) remains unchanged.

The Lagrangian (\ref{LpotSO7}) after redefinition (\ref{defu}), (\ref{5DNewFields}) takes the remarkably compact form
\bea
{\cal L}_{\rm pot} =
e^{-3\Phi} \left(
3\,\partial_u\Phi \,\partial_u \Phi 
-2\,\partial_u\Phi_1 \,\partial_u \Phi_1
-2\,e^{-4\Phi_1}\,\partial_u{\cal X}_2 \,\partial_u {\cal X}_2
\right)
+\frac{15}{2}\,e^{-\Phi-2\,\Phi_1}
\;.
\label{L5D}
\eea
As a result, the action given by (\ref{LkinSO7}) and (\ref{L5D}) (upon temporarily relaxing the constraints (\ref{gA0})) can be written in 
manifestly five-dimensional form as
\bea
S=\int d^4x\,du\,
\sqrt{|G_{(5)}|}\;\Big( R^{(5)} 
-2\,\partial_{\hat\mu}\Phi_1 \,\partial^{\hat\mu} \Phi_1
-2\,e^{-4\Phi_1}\,\partial_{\hat\mu}{\cal X}_2 \,\partial^{\hat\mu} {\cal X}_2
+\frac{15}{2}\,e^{-2\,\Phi_1}
\Big)
\;,
\label{uplift5DSO7}
\eea
with $\{x^{\hat\mu}\}=\{x^\mu, u\}$, $\hat\mu=0, \dots, 4$\,. This is the action of $D=5$ gravity coupled to 
an ${\rm SL}(2)/{\rm SO}(2)$ sigma model. It results from the consistent truncation of $D=11$ supergravity 
to the ${\rm SO}(6)$-singlet modes around the round six-sphere $S^6={\rm SO}(7)/{\rm SO}(6)$.
Splitting the 5D metric in the standard Kaluza-Klein fashion
\bea
G_{\hat\mu\hat\nu} &=&
\begin{pmatrix}
e^{-\Phi} g_{\mu\nu} + e^{2\Phi} A_\mu A_\nu & e^{2\Phi} A_\mu \\
e^{2\Phi} A_\mu  & e^{2\Phi}
\end{pmatrix}
\;,
\eea
reinstating the truncation (\ref{gA0}), 
together with inverting the change of coordinates (\ref{defu}) and fields (\ref{5DNewFields}),
the Lagrangian (\ref{uplift5DSO7}) then yields back the ExFT Lagrangian 
(\ref{LkinSO7}), (\ref{LpotSO7}).

As seen above, solutions of type AdS$_4\times \Sigma_7$ correspond to regular boundary behaviour 
(at the endpoints of the interval $\theta=0, \theta=\pi$) for the fields $\{\phi, \phi_1, \chi_2\}$, 
which in turn will correspond to divergent boundary behaviour
of $\{\Phi, \Phi_1, {\cal X}_2\}$ at the endpoints of the interval $u\in[-1,1]$.
In other words, in the $D=5$ reformulation (\ref{uplift5DSO7}) of this truncation, we need to identify 
particular singular solutions in order to describe a regular AdS$_4\times \Sigma_7$ geometry.

\subsubsection{Field equations}

The reformulation (\ref{uplift5DSO7}) of the ${\rm SO}(7)$-singlet sector of $D=11$ supergravity
allows to quickly derive and further simplify the equations of motion.
Let us first note that the global symmetry associated with the conserved charge $V$ from (\ref{V})
is nothing but the invariance of (\ref{L5D}) under translations in $u$, i.e.\ corresponds to the conserved  
`energy' of this one-dimensional Lagrangian. 
Moreover, the field equations following from the Lagrangian (\ref{L5D})  imply that
\begin{equation}
e^{-3\Phi-4\Phi_1}\,\partial_u {\cal X}_2 = F \;,
\label{FSO7}
\end{equation}
with some constant $F$, corresponding to equation (\ref{dCc}) in the previous variables. 
In consequence, this equation can be used to eliminate ${\cal X}_2$ from the Lagrangian and arrive at
\bea
{\cal L}_{\rm pot, red} =
e^{-3\Phi} \left(
3\,(\partial_u \Phi)^2
-2\,(\partial_u \Phi_1)^2
\right)
+ 2\, e^{3\Phi+4\Phi_1}\,F^2
+\frac{15}{2}\,e^{-\Phi-2\,\Phi_1}
\;.
\label{L5D2}
\eea
The conserved charge $V$ from (\ref{V}) then is simply given by
\bea
V = 
e^{-3\Phi} \left(
3\,(\partial_u \Phi)^2
-2\,(\partial_u \Phi_1)^2
\right)
- 2\, e^{3\Phi+4\Phi_1}\,F^2
-\frac{15}{2}\,e^{-\Phi-2\,\Phi_1}
\;.
\label{VD5}
\eea

Let us spell out the field equations obtained from (\ref{L5D2}), however re-expressed in terms of the 
original fields $\phi$ and $\Delta$, in order to better illustrate the boundary asymptotics
\bea
0&\!=\!&
720\,  \Delta^{-3/2} \, e^{\frac{7 \phi}{2}}
-720\, \cos^2\theta
-36  \,\sin (2\theta ) \left( \Delta^{-1} \partial_\theta\Delta
-5\, \partial_\theta\phi \right)
\nonumber\\
&&{}
+\sin^2\theta\left(
-48 \Delta^{-1} \partial^2_\theta\Delta
-45 \,(\partial_\theta\phi)^2   
+90 \,\Delta^{-1}\partial_\theta\Delta \partial_\theta\phi
+75 \,\Delta^{-2} (\partial_\theta\Delta)^2
-320\,F^2 \Delta^3 \,e^{3 \phi}
\right)
,\;\;\;\;
\nonumber\\[1ex]
0 &\!=\!& 
80 \left(\Delta^{-3/2} e^{\frac{7 \phi}{2}}-1\right)
+12 \,\sin (2\theta ) \left(- 3\, \Delta^{-1}\partial_\theta \Delta +7 \, \partial_\theta\phi \right)
\nonumber\\
&&{}
+ \sin^2\theta   \left(16\, \partial^2_\theta\phi -33\, (\partial_\theta\phi)^2+18 \Delta^{-1}  \partial_\theta\Delta\, \partial_\theta\phi-9 \, \Delta^{-2}\,(\partial_\theta\Delta)^2
-64 \,F^2 \Delta^3\, e^{3 \phi}+144\right)
.
\label{systEQ}
\eea
In these fields, the conserved charge (\ref{VD5}) takes the form
\bea
V&=&
\frac1{32}\,
e^{-3 \phi} \,\Big(
18 \Delta^{-1} \partial_\theta\Delta \left(\partial_\theta\phi-4 \cot (\theta )\right)
-9 \,\Delta^{-2}\,\partial_\theta\Delta^2
+15\,\left(\partial_\theta\phi-4 \cot (\theta )\right)^2\Big)
\nonumber\\
&&{}
-\frac{15}{2} \,\Delta^{-3/2}\, e^{\phi/2}\, {\rm csc}^2\theta
-2 \,F^2\, \Delta^3
~=~
-\frac{3}{\ell_4^2}
\;.
\label{VD11}
\eea
Expanding the system of ordinary differential equations (\ref{systEQ}) near the boundaries of the interval $\theta\in[0,\pi]$
exhibits the singularities. Imposing regularity of the solutions $\phi$, $\Delta$ at the boundary requires both to be even functions in $\theta$
with the lowest coefficients in their Taylor expansion restricted by
\begin{equation}
\Delta|_{\theta=0} = e^{7\phi/3}|_{\theta=0}
\;,\qquad
\partial^2_\theta \Delta|_{\theta=0} = \frac1{33}\left[e^{7\phi/3}(36+81\,\partial^2_\theta\phi -16\,e^{10\phi}\,F^2) \right]_{\theta=0}
\;.
\label{eq:singSO7}
\end{equation}
A solution regular at $\theta=0$ thus is determined by two integration
constants, which may be chosen to be $\phi(0)$ and $\partial^2_\theta\phi|_{\theta=0}$\,.
A generic solution of this type will be singular at the other end $\theta=\pi$ of the interval.
Inducing regularity at both endpoints of the interval thus reduces the set of solutions to a 
discrete set.

Before analysing possible regular solutions in more detail, let us note that
we may recover two analytic solutions of the system (\ref{systEQ}),
both corresponding to known solutions living within the consistent truncation to ${\cal N}=8$ supergravity  
\cite{deWit:1982bul}
\bea
{\rm SO}(8) &:& \phi=0\;,\quad
\Delta=1\;,\quad
F=\frac{3}{2}\;,\quad
\ell_4=\frac12\;,
\nonumber\\[1ex]
{\rm SO}(7)_+ &:&
\phi=-\frac14\,{\rm ln}\,5\;,\quad
\Delta=\frac{5^{1/12}}{(3+2\,{\rm cos}(2\theta))^{2/3}}\;,\quad
F=\frac{5^{3/4}}{2}\;,\quad
\ell_4=\frac{3^{1/2}}{2\cdot 5^{3/8}}\;.
\eea
The first solution is the round sphere $S^7$, the second one corresponds to the ${\rm SO}(7)$-squashed 
$S^7$ found in \cite{deWit:1984va}.

\subsection{${\rm G}_2$ truncation and uplift to $D=11$}

Having described in detail the simplification of the consistent truncation to ${\rm SO}(7)$-singlets,
eventually described by the simple $D=5$ Lagrangian (\ref{L5D2}), we can now extend the discussion
to the full sector of ${\rm G}_2$-singlets.
Recall, that the Lagrangian obtained from ExFT is given by (\ref{LpotG2}), (\ref{Dtheta}) in terms of
six scalar fields parametrising the coset space (\ref{eq:cosetCEKG2}).
Following the previous discussion, we apply the coordinate transformation (\ref{defu}) together with a field redefinition 
\begin{equation}
\left\{\phi,\chi, \phi_1, \chi_{1a}, \chi_{1b}, \chi_2 
\right\}
\longrightarrow
\left\{\Phi,{\cal X}, \Phi_1, {\cal X}_A, {\cal X}_B, {\cal X}_2 
\right\},
\label{phiphiG2}
\end{equation}
by a non-linear $\theta$-dependent ${\rm SL}(2)\times {\rm SU}(2,1)$ transformation,
generalising (\ref{5DNewFields}). After this redefinition, 
the action (\ref{LpotG2}) takes the compact form
\bea
{\cal L}_{\rm pot} &=&
{3}\,e^{-3 \Phi }(\partial_u \Phi )^2
-2\,e^{-3 \Phi }\,(\partial_u\Phi_1)^2
-2\,e^{-3 \Phi -4\Phi_1}\,
M_{mn}\,D_u {\cal X}^m D_u {\cal X}^n
\nonumber\\
&&{}
+\frac{15}2\,e^{ -\Phi -2\Phi_1}-6\,e^{ -\Phi -4\Phi_1}\,{\cal X}_A^2\;,
\label{LpotG25D0}
\eea
with $\left\{ {\cal X}^m\right\}=\left\{{\cal X}_A, {\cal X}_B, {\cal X}_2\right\}$, the matrix 
\bea
M_{mn} =
\begin{psmallmatrix}
 e^{2\Phi_1}+{\cal X}_B^2 & -{\cal X}_A{\cal X}_B&-{\cal X}_B \\[.5ex]
-{\cal X}_A{\cal X}_B & e^{2\Phi_1}+{\cal X}_A^2 &{\cal X}_A \\[.5ex]
-{\cal X}_B & {\cal X}_A& 1
\end{psmallmatrix}
\;,
\label{MijN}
\eea
and the `covariant' derivatives $D_u$ defined as
\begin{equation}
D_u {\cal X}_A  =
\partial_u{\cal X}_A
  \;,\quad
D_u{\cal X}_B  =\partial_u {\cal X}_B
-3 \,{\cal X}
  \;,\quad
  D_u{\cal X}_2  =
\partial_u {\cal X}_{2} -3\,{\cal X}{\cal X}_A
  \;. 
  \label{Dthetau}
\end{equation}
The kinetic term (\ref{LkinG2}) is invariant under the transformation (\ref{phiphiG2}).
The full truncation to ${\rm G}_2$-singlets can then be written in 
manifestly five-dimensional form upon extending (\ref{uplift5DSO7}) to
\bea
S&=&\int d^4x\,du\,
\sqrt{|G_{(5)}|}\;\Big( {\cal L}_{{\rm 5D, min}}
-2\,\partial_{\hat\mu}\Phi_1 \,\partial^{\hat\mu} \Phi_1
-2\,e^{-4\Phi_1} M_{ij}\,D_{\hat\mu} {\cal X}^i D^{\hat\mu} {\cal X}^j
\nonumber\\
&&{}
\qquad\qquad\qquad\qquad
\;\;+\frac{15}2\,e^{-2\Phi_1}-6\,e^{-4\Phi_1}\,{\cal X}_A^2\,
\Big)
\;,
\label{uplift55DG2}
\eea
with
\begin{equation}
D_{\hat\mu} {\cal X}_A  =
\partial_{\hat\mu}{\cal X}_A
  \;,\quad
D_{\hat\mu}{\cal X}_B  =\partial_{\hat\mu} {\cal X}_B
-3 \,A_{\hat\mu} 
  \;,\quad
  D_{\hat\mu}{\cal X}_2  =
\partial_{\hat\mu} {\cal X}_{2} -3\,A_{\hat\mu} {\cal X}_A
  \;,
  \label{Dthetau5}
\end{equation}
with $\{x^{\hat\mu}\}=\{x^\mu, u\}$, $\hat\mu=0, \dots, 4$\,. The first term in (\ref{Dthetau5}) is the bosonic sector of
minimal supergravity in $D=5$, i.e.\ describes a vector field $A_\mu$ with $D=5$ Chern-Simons term coupled to $D=5$ gravity.
The remaining part of (\ref{Dthetau5}) describes the coupling to one hypermultiplet
with target space ${\rm SU}(2,1)/{{\rm U}(1)}$ and gauging of a shift isometry according to (\ref{Dthetau})
(upon identification of ${\cal X}$ with the fifth component $A_u$ of the gauge field).
The Lagrangian (\ref{Dthetau5}) results from the consistent truncation of $D=11$ supergravity to 
the G$_2$-singlets around the six-sphere $S^6={{\rm G}_{2}}/{\SU{3}}$. Its form is consistent with the fact
that this truncation retains one $D=5$ gravitino, thus describes the bosonic sector of a $D=5$, ${\cal N}=1$ supergravity.

In the search for AdS$_4$ solutions, we again impose (\ref{gA0}) and require scalar fields to be constant in AdS$_4$ spacetime,
such that the system is described by the ordinary differential equations obtained from variation of the Lagrangian (\ref{LpotG25D0}).
Variation w.r.t.\ ${\cal X}_2$ implies that
\begin{equation}
e^{-3\Phi-4\Phi_1}\left(
\partial_u{\cal X}_2+{\cal X}_A\,\partial_u{\cal X}_B- {\cal X}_B \partial_u{\cal X}_A-6{\cal X}_A{\cal X}
\right) = F
\;,
\label{defF}
\end{equation}
with some constant $F$, generalising equation (\ref{FSO7}). Moreover, the field ${\cal X}$ 
appears only algebraically in (\ref{LpotG25D0}), entering the covariant derivatives (\ref{Dthetau}). It can thus be eliminated by
its own field equation
\begin{equation}
3\,{\cal X} = \partial_u {\cal X}_B+2\,e^{3\Phi+2\Phi_1}\,F{\cal X}_A
\;,
\end{equation}
where we have already used (\ref{defF}) for simplification. Upon integrating out ${\cal X}$ and ${\cal X}_2$,
we are thus left with the one-dimensional Lagrangian
\bea
{\cal L}_{\rm pot, red} &=&
{3}\,e^{-3 \Phi }(\partial_u \Phi )^2
-2\,e^{-3 \Phi }\,(\partial_u\Phi_1)^2
-2e^{-3 \Phi -4\Phi_1}\,(\partial_u{\cal X}_A )^2
\nonumber\\
&&{}
+\frac{15}2\,e^{ -\Phi -2\Phi_1}-6\,e^{ -\Phi -4\Phi_1}\,{\cal X}_A^2
+2\,e^{3\Phi+4\Phi_1}\,F^2\,
+8\,e^{3\Phi+2\Phi_1}\,F^2 {\cal X}_A^2\;,
\label{LpotG25D}
\eea
describing all the equations that define an AdS$_4$ solution. In particular, we note that the field ${\cal X}_B$ has also 
disappeared from the Lagrangian, such that we are left with a system of three scalar fields. 
This is a remnant of the gauge freedom and the Higgs effect in the full $D=11$ theory.

Before analysing the field equations and their solutions, let us first spell out the uplift of the model to $D=11$ dimensions.
For the $D=11$ metric, we have given the result in  (\ref{metric11DG2}) above
\begin{equation}
ds^2 = \Delta^{-1}\,ds_{(4)}^2
+ e^{3\phi}\,\Delta^{-1}\,d\theta^2
+e^{-\frac{\phi}{2}}\,{\rm sin}^2\theta\,\Delta^{1/2}\,ds_{S^6}^2
\;,
\label{metric11DG2-2}
\end{equation}
where $\phi$ and $\Delta$ are related to the fields of (\ref{LpotG25D}) via (\ref{5DNewFields})
\begin{equation}
e^{\Phi} = \rho^{4/3}\,e^{\phi}
\;,\qquad
e^{\Phi_1} = \rho^{-1}\,e^{-3\phi/4} \,\Delta^{3/4}
\;.
\label{eq:newv1}
\end{equation}
In particular, the determinant of the metric on the internal space is given by
\bea
{\rm det} \,g_{(7)} &=&
\rho^{-4}\,\Delta^{2}\,{\rm det} \,g_{S^6}
\;.
\eea
Similarly, one obtains the uplift for the $D=11$ three-form $C_{(3)}$ and its field strength $F_{(4)}$ as
\bea
F_{(4)} &=& 4\,F\, \omega_{(4)}+dC_{(3)}
\;,\nonumber\\
C_{(3)}&=&\frac16\,\sin^4\theta\,A\left(
c_{ijn}c_{kmn}\,y^m dy^i dy^j dy^k 
-{4F}\, e^{3\phi/2}\,\Delta^{3/2}\,c_{ijk}\,y^i\,dy^jdy^k d\theta\right)
\;,
\label{11DfieldsG2}
\eea
after redefining
\begin{equation}
{\cal X}_A = \rho^{-4/3}\,A
\;,
\label{eq:newv2}
\end{equation}
and where $F$ is the constant introduced in (\ref{defF}).
The result is given in terms of the embedding coordinates $y^i$, $i=1, \dots 7$, of the round $S^6$, $y^iy^i=1$, 
while $c_{ijk}$ is the unique totally antisymmetric cubic G$_2$-invariant tensor, normalised as $c_{ijk}c^{ijk}=42$\,.
$\omega_{(4)}$ is the AdS$_4$ volume form.

In turn, the ansatz (\ref{metric11DG2-2}), (\ref{11DfieldsG2}) is the most general G$_2$-invariant 
ansatz for an AdS$_4\times \Sigma_7$ solution of $D=11$ supergravity, after gauge fixing
of the $D=11$ tensor gauge symmetries.
The metric (\ref{metric11DG2-2}) still has the full ${\rm SO}(7)$ isometry group, 
which is broken to G$_2$ by the three-form $C_{(3)}$.
An early analysis of G$_2$-invariant compactifications \cite{Gunaydin:1983mi}
was restricted to a constant warp factor $\Delta=1$, which leaves the system with only two solutions, 
denoted as ${\rm SO}(8)$ and ${\rm SO}(7)_-$ below. The subsequent analysis of \cite{deWit:1984nz}
allowed for a warp factor, but was restricted to solutions that live within the consistent truncation to ${\cal N}=8$ supergravity. In particular,
this implies $\phi={\rm const}$, and leaves the system with four solutions, given in the next subsection.
Relaxing these restrictions, we will find new numerical solutions in subsection~\ref{subsec:numerics} below.

\subsection{Field equations and analytic solutions}

For the further analysis, we spell out the equations of motion derived from variation of (\ref{LpotG25D})
in terms of the coordinate $\theta$ and the fields (\ref{eq:newv1}), (\ref{eq:newv2})
which directly feature in the expressions for the $D=11$ fields.
Explicitly, these equations are given by
\bea
0&=&
\sin^2\theta\left(
-48 \,\Delta^{-1} \partial^2_\theta\Delta
-45 \,(\partial_\theta\phi)^2   
+90 \,\Delta^{-1}\partial_\theta\Delta \partial_\theta\phi
+75 \,\Delta^{-2} (\partial_\theta\Delta)^2
-320\,F^2 \,\Delta^3 \,e^{3 \phi}
\right)
\nonumber\\
&&{}
+720\,  \Delta^{-3/2} \, e^{\frac{7 \phi}{2}}
-720\, \cos^2\theta
-36  \,\sin (2\theta ) \left( \Delta^{-1} \partial_\theta\Delta
-5\, \partial_\theta\phi \right)
\nonumber\\
&&{}
-16\,e^{3\phi/2}\,\Delta^{-3}\,\sin^2\theta\left(
\Delta^{3/2}\left(4A\cos\theta+\partial_\theta A\sin\theta\right)^2
+84\,e^{7\phi/2}\,A^2 +16\,e^{3\phi}\,F^2\,\Delta^{9/2}A^2\,\sin^2\theta
\right)
,\nonumber\\[1ex]
0 &=& 
\sin^2\theta   \left(16\, \partial^2_\theta\phi -33\, (\partial_\theta\phi)^2+18 \,\Delta^{-1}  \partial_\theta\Delta\, \partial_\theta\phi-9 \, \Delta^{-2}\,(\partial_\theta\Delta)^2
-64 \,F^2\, \Delta^3\, e^{3 \phi}+144\right)
\nonumber\\
&&{}
+80 \left(\Delta^{-3/2} e^{\frac{7 \phi}{2}}-1\right)
+12 \,\sin (2\theta ) \left(- 3\, \Delta^{-1}\partial_\theta \Delta +7 \, \partial_\theta\phi \right)
\nonumber\\
&&{}
-16\,e^{3\phi/2}\,\Delta^{-3}\,\sin^2\theta\left(
\Delta^{3/2}\left(4A\cos\theta+\partial_\theta A\sin\theta\right)^2
+4\,e^{7\phi/2}\,A^2 +16\,e^{3\phi}\,F^2\,\Delta^{9/2}A^2\,\sin^2\theta
\right)
,\nonumber\\[1ex]
0 &=& 
\sin^2\theta\left(
-2\,\partial_\theta^2A+3\, \partial_\theta A\,(\partial_\theta\phi+ \Delta^{-1}\partial_\theta\Delta) 
+32\,A\left(1-e^{3\phi}\,F^2\,\Delta^3\right)
\right)
\nonumber\\
&&{}
+\sin(2\theta)\left(
6\,A\,(\partial_\theta\phi+ \Delta^{-1}\partial_\theta\Delta)-8\,\partial_\theta A
\right)
+24\,A\,(e^{7\phi/2}\,\Delta^{-3/2} -1)
.
\label{systEQG2}
\eea
The system admits a conserved charge, corresponding to the invariance of the system (\ref{LpotG25D}) 
under translations in $u$, originally given in (\ref{V}) and related to the  AdS$_4$ radius by (\ref{ell4}).
In terms of the fields $\{\phi, \Delta, A\}$, it takes the explicit form
\bea
V&=&
\frac1{32}\,
e^{-3 \phi} \,\Big(
18 \Delta^{-1} \partial_\theta\Delta \left(\partial_\theta\phi-4 \cot (\theta )\right)
-9 \,\Delta^{-2}\,\partial_\theta\Delta^2
+15\,\left(\partial_\theta\phi-4 \cot (\theta )\right)^2\Big)
\nonumber\\
&&{}
-\frac{15}{2} \,\Delta^{-3/2}\, e^{\phi/2}\, {\rm csc}^2\theta
-2 \,F^2\, \Delta^3
-\frac12\,e^{-3\phi/2}\,\Delta^{-3/2}\,\left(4A\cos\theta+\partial_\theta A\sin\theta\right)^2
\nonumber\\
&&{}
+2\left(
3\,e^{2\phi}\,\Delta^{-3}-4\,e^{3\phi/2}\,\Delta^{3/2}\,F^2 \sin^2\theta\right) A^2
~=~
-\frac{3}{\ell_4^2}
\;,
\label{VD11G2}
\eea
generalising (\ref{VD11}) to the full G$_2$ truncation. One may check explicitly, that $V$ is conserved, $\partial_\theta V=0$,
as a consequence of the equations (\ref{systEQG2}).

Equations (\ref{systEQG2}) are invariant under the scaling symmetry
\begin{equation}
e^\phi\rightarrow \lambda^3\,e^\phi\;,\quad
\Delta \rightarrow \lambda^7\,\Delta\;,\quad
A\rightarrow \lambda^3\, A\;,\quad
F\rightarrow \lambda^{-15}\,F
\;,
\qquad
\lambda\in\mathbb{R}^*
\;,
\label{scaling}
\end{equation}
with constant $\lambda$. This is the trombone symmetry of $D=11$ supergravity \cite{Cremmer:1997xj},
under which the AdS$_4$ radius $\ell_4$ (\ref{VD11G2}) scales as
\begin{equation}
\ell_4\rightarrow \lambda^{9/2}\,\ell_4 
\;.
\end{equation}
For the subsequent numerical analysis, we fix this scaling symmetry (\ref{scaling}) to set
the constant $F$ from (\ref{defF}) to 
\begin{equation}
F=\frac32\,.
\label{F32}
\end{equation}

All previously known solutions to the equations (\ref{systEQG2}) are analytic, have constant $\phi$, and live within the
consistent truncation to ${\cal N}=8$ supergravity \cite{deWit:1982bul}. They correspond to the 
four G$_2$-invariant extremal points of the scalar potential \cite{Warner:1983vz}.
In our conventions, in particular after having fixed the scaling symmetry by (\ref{F32}), they take the form
\bea
{\rm SO}(8) &:&
e^\phi=1\;,\;\;
\Delta=1\;,\;\;
A=0
\;,\nonumber\\
{\rm SO}(7)_+ &:&
e^\phi=3^{-1/5}\,5^{-1/10}\;,\;\;
\Delta=5^{13/30}\,3^{-7/15}\,(1+4\cos^2\theta)^{-2/3}\;,\;\;
A=0
\;,\nonumber\\
{\rm SO}(7)_- &:&
e^\phi=2^{1/5}\,3^{-1/5}\;,\;\;
\Delta=2^{7/15}\,3^{-7/15}\;,\;\;
A=2^{-4/5}\,3^{-1/5}
\;,\nonumber\\
{\rm G}_2 &:& 
e^\phi=3^{-3/10}\,,\;\;
\Delta=3^{-1/30}\,(1+2\cos^2\theta)^{-2/3}\,,\;\;
A=3^{1/5}\,5^{-1/2}\,(1+2\cos^2\theta)
\;.
\;\;
\label{sol_analytic}
\eea

\subsection{New numerical solutions}
\label{subsec:numerics}

In the rest of this paper, we will discuss the equations of motion (\ref{systEQG2}) and their solutions by numerical analysis.
To this end, we go back to coordinate $w$ from (\ref{wt}).
As we have already discussed for the SO(7) subsector, c.f.\ (\ref{eq:singSO7}) above, 
 the system of second order differential equations (\ref{systEQG2}) is singular at the boundary of the interval $\theta\in[0,\pi]$,
 i.e.\ $w=\pm1$\,.
As a consequence, for a regular solution only three of the (a priori six) initial conditions can be chosen freely at $w=1$, 
and we choose these to be
\begin{equation}
\phi(1)\equiv \mathfrak{q}\;,\quad
\phi'(1)\equiv \mathfrak{p}\;,\quad
A(1)\equiv \mathfrak{a}
\;.
\label{initial}
\end{equation}
Throughout this section, primes refer to derivatives w.r.t.\ $w$: $\phi'=\partial_w\phi$, etc..
Regularity of the solution at $w=1$ then determines the next coefficients in the respective Taylor expansions
\bea
&&
\Delta(1)=e^{7\mathfrak{q}/3}\;,\quad
\Delta'(1)=\frac{1}{33} e^{\mathfrak{q}/3} \left(80  \mathfrak{a}^2+16\,F^2\,e^{12 \mathfrak{q}}
+9 e^{2\mathfrak{q}} \left(9 \mathfrak{p}-4\right)\right)
\;,\nonumber\\
&&
A'(1)\to \frac{4}{99} \, \mathfrak{a}\, e^{-2\mathfrak{q}} \left(-20  \mathfrak{a}^2+40 \,F^2\, e^{12 \mathfrak{q}}
+e^{2 \mathfrak{q}} \left(54 \mathfrak{p}-35\right)\right)
\;,
\eea
and similarly, all higher coefficients in the Taylor expansion are fixed by expanding the equations (\ref{systEQG2}).
For generic choice of the boundary conditions (\ref{initial}), the solution will however be singular at the other endpoint $w=-1$ 
of the interval, or even diverge before reaching the endpoint.
Further imposing regularity at the opposite boundary $w=-1$ thus imposes three (highly non-linear) relations among the 
parameters (\ref{initial}) such that a naive counting argument indicates that the system allows for only a discrete set of regular solutions.
Indeed, that is what we observe in the following. 

Let us also note that the cosmological constant (\ref{VD11G2}) is given as a function of the boundary conditions (\ref{initial}) as
\begin{equation}
\ell_4^2 = \frac{66 \,e^{5\mathfrak{q}}}{180\mathfrak{a}^2+80 \,F^2\, e^{12 \mathfrak{q}}-21 e^{2 \mathfrak{q}}
   \left(9\, \mathfrak{p}-4\right)}
   \;.
\end{equation}

In the following numerical analysis, we will separate the cases $A=0$, which amounts to truncating to the subsector
of SO(7)-singlets discussed in section~\ref{sec:SO7}, and $A\not=0$. In total, we find three new numerical solutions on top of the 
known analytic solutions (\ref{sol_analytic}). Our findings are summarised in Table~\ref{tab:solutions}.

\begin{table}[bt]
\centering
\begin{tabular}{c|| c|c|c|c|l}
solution &$\mathfrak{q}$&$\mathfrak{p}$&$\mathfrak{a}$ & $\ell_4$ & comments\\
\hline
SO(8) & 0 & 0 & 0 & $0.500000$ & \cite{Duff:1983gq}, ${\cal N}=8$, round $S^7$ \\
\hline
SO(7)$_-$ & $-0.0810930$ &  0 & $0.461054$ & $0.497590$ & \cite{Englert:1982vs}, `parallelised' $S^7$
\\
\hline
SO(7)$_+$ & $-0.380666$ & 0 &0  & $0.489270 $ & \cite{deWit:1984va}
\\
\hline
G$_2$ & $-0.329584$ &0 & $0.185703$ & $0.489049$ & \cite{deWit:1984nz}, ${\cal N}=1$ \\
\hline\hline
${\rm SO}(7)'$ & $-0.250533$ & $-0.137962$ & 0 & $0.499467$&  new, preserves SO(7)\\
\hline
G$_2'$ & $-0.202438$ & $-0.105189$ &  $0.225857$ & $0.504244$ & new \\
\hline
G$_2''$ & $0.0544548$ & $0.892275$ & $ 0.658650$ & $0.512668$ & 
new \\
\hline
\end{tabular}
\caption{List of regular G$_2$-invariant solutions of the system (\ref{systEQG2}). The first four solutions live within the
consistent truncation to ${\cal N}=8$ supergravity \cite{deWit:1982bul} and correspond to the 
four G$_2$-invariant extremal points of the scalar potential \cite{Warner:1983vz}.
They have been known before and can be given in analytic form (\ref{sol_analytic}).
The three last lines are the new numerical solutions.
 All digits displayed are within the numerical accuracy.}
\label{tab:solutions}
\end{table}

\subsubsection{$A=0$}
\label{subsec:A0}

We first discuss the subsector with $A=0$, which is a consistent truncation of the 
system (\ref{systEQG2}), corresponding 
to the subsector of SO(7)-singlets discussed in section~\ref{sec:SO7}.
We then scan the two-dimensional parameter space of initial conditions $\{\mathfrak{q}, \mathfrak{p}\}$
for solutions regular at $w=1$,
searching for solutions regular throughout the interval $w\in[-1,1]$.
While regularity at $w=-1$ is hard to control, we note that the problem can be simplified for
even solutions satisfying
\begin{equation}
\phi(-w)=\phi(w)\,,\quad
\Delta(-w)=\Delta(w)\,,
\label{eq:even}
\end{equation}
corresponding to a $\mathbb{Z}_2$-symmetry of the system (\ref{systEQG2}). 
Starting from a solution regular at $w=1$, we search for initial conditions such that the solution satisfies
\bea
\phi'(0)=0=\Delta'(0)
\;,
\label{eq:phiprime}
\eea
at $w=0$. This implies the symmetry (\ref{eq:even}) and regularity at the other endpoint $w=-1$ becomes a simple
consequence of this symmetry. The conditions (\ref{eq:phiprime}) can be straightforwardly
implemented into a numerical search. For a regular solution to exist, however,
both conditions (\ref{eq:phiprime}) must hold exactly, not just approximately.

\begin{figure}[tb]
\center
\includegraphics[scale=.24]{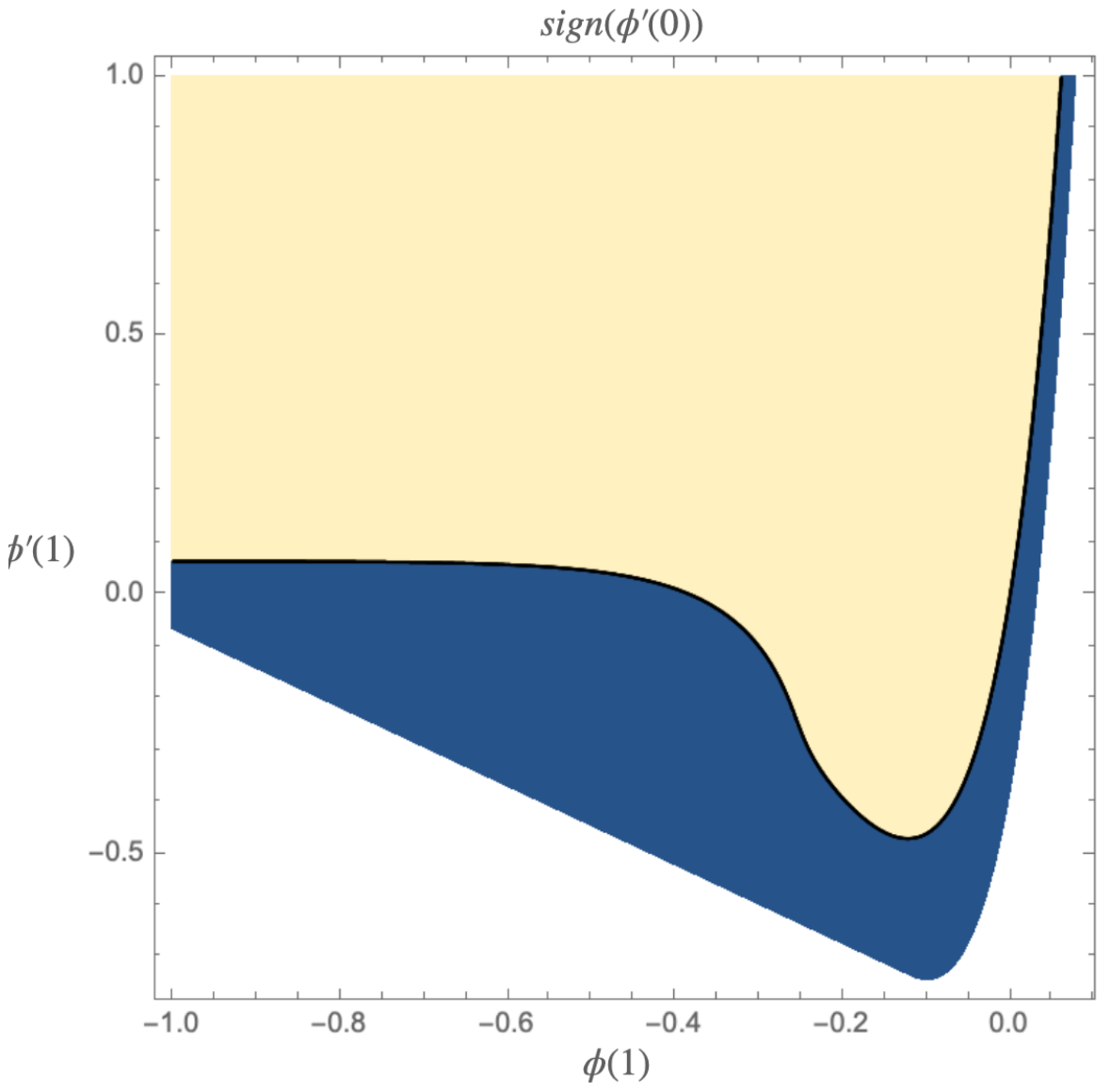}
\includegraphics[scale=.24]{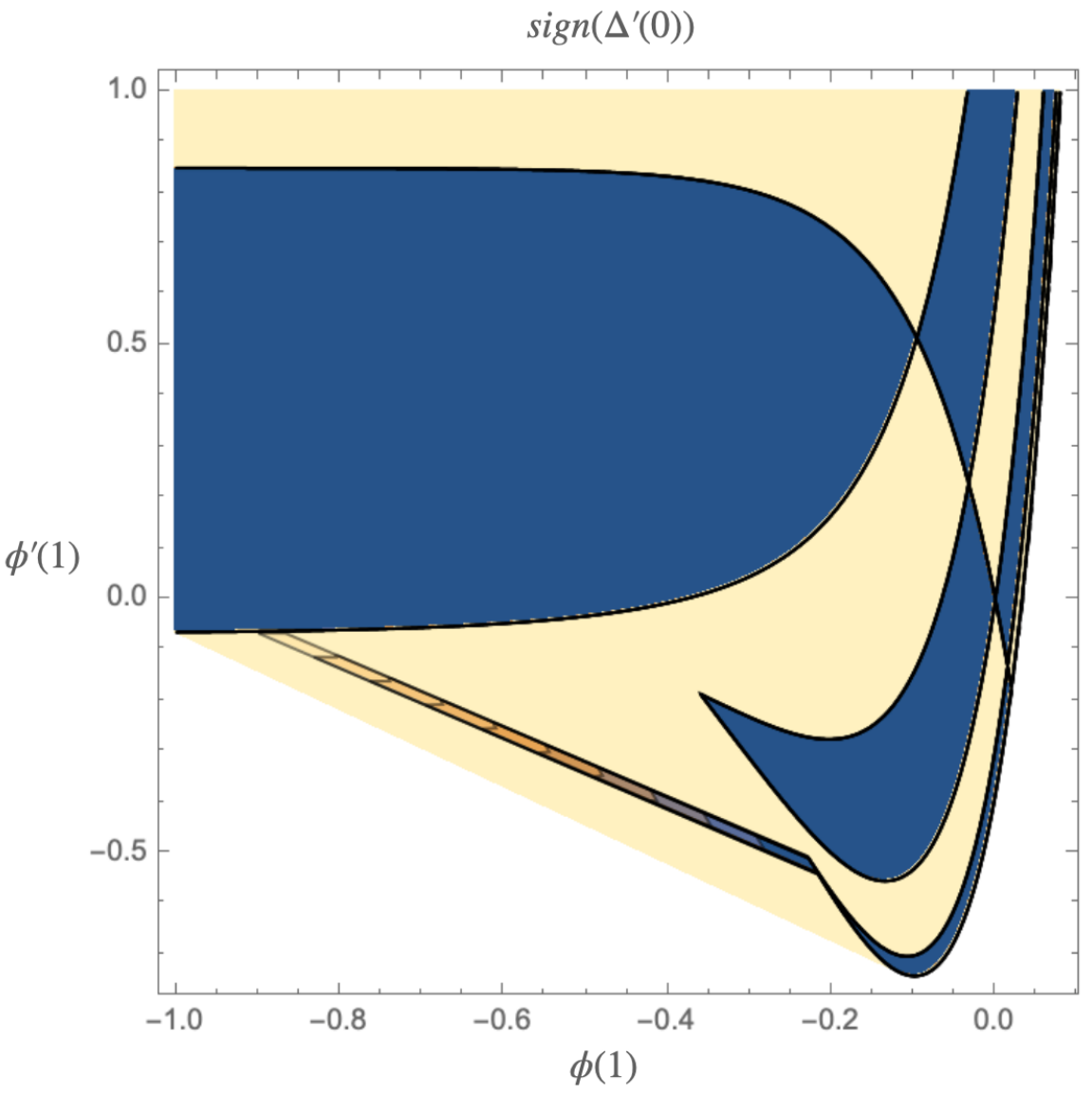}
\quad
\includegraphics[scale=.30]{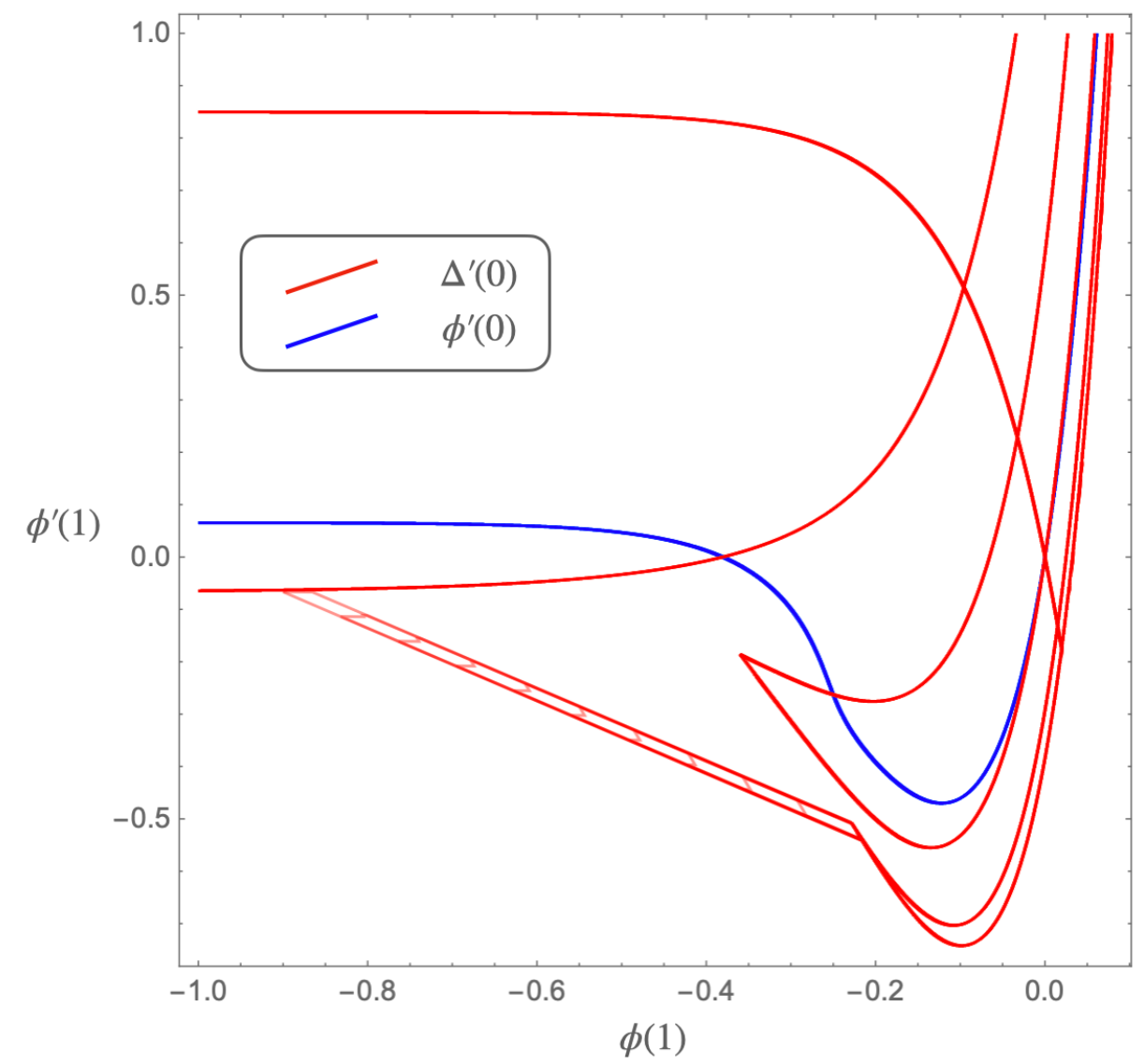}
\caption{
Initial values for the regular solutions in the SO(7) system with vanishing $A=0$.
The first two plots show the regions in the two-dimensional parameter space $(\mathfrak{q}, \mathfrak{p})$ in which the 
signs of $\phi'(0)$ and $\Delta'(0)$ are positive (yellow) and negative (blue), respectively.
The third plot extracts the lines of vanishing $\phi'(0)$ (blue) and vanishing $\Delta'(0)$ (red).
The three intersection points of the red and blue lines in this plot correspond to the solutions ${\rm SO}(8)$,  ${\rm SO}(7)_+$,
and  ${\rm SO}(7)'$ from Table~\ref{tab:solutions}.
}
\label{fig:linesE}
\end{figure}

\begin{figure}[bt]
\center
\includegraphics[scale=.70]{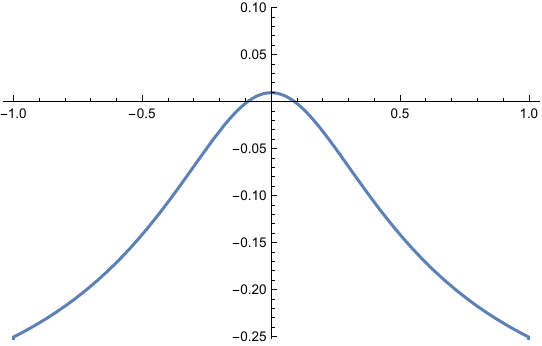}\qquad
\includegraphics[scale=.70]{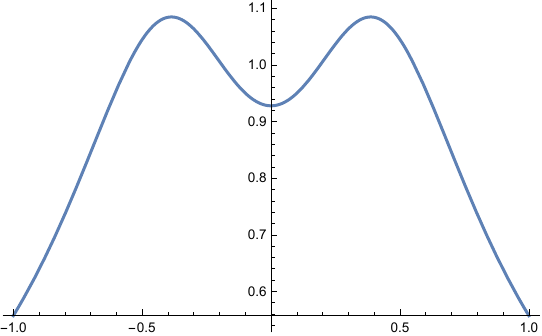}
\caption{New solution SO(7)': fields $\phi$ and $\Delta$ as functions of $w$.}
\label{fig:so7'}
\end{figure}

To this end, we first identify the lines in the parameter space of initial conditions $(\mathfrak{q}, \mathfrak{p})$, 
along which $\phi'(0)$ and $\Delta'(0)$ vanish separately. Even before optimizing the numerical accuracy, we can infer the
existence of such lines by identifying the regions in parameter space in which the signs of $\phi'(0)$ and $\Delta'(0)$ 
are positive and negative, respectively.
Concretely, we depict in the first plot of Figure~\ref{fig:linesE} the yellow region in which $\phi'(0)$ is positive and
the blue region in which $\phi'(0)$ is negative. The interface between the two regions then defines a line along which $\phi'(0)$
vanishes. In the second plot of Figure~\ref{fig:linesE}, we depict the analogous information for $\Delta'(0)$. We then extract the lines of vanishing $\phi'(0)$ (blue) and vanishing $\Delta'(0)$ (red) in the third plot, which shows the existence of three intersection points at which both conditions (\ref{eq:phiprime}) are satisfied.\footnote{A better resolution of the hatched zone in the second plot of Figure~\ref{fig:linesE} would require to improve the numerical accuracy. However, the third plot shows that this region is not close to any blue line, thus irrelevant for the search of solutions.}
Once, we have established the existence of such intersection points, we can work on improving the numerical accuracy of the corresponding solutions.
Two of these points correspond to the known ${\rm SO}(8)$,  
and ${\rm SO}(7)_+$ solutions from (\ref{sol_analytic}), the third one represents a new ${\rm SO}(7)$-invariant solution, which we will denote as ${\rm SO}(7)'$. We plot the fields $\phi$ and $\Delta$ for the new numerical solution in Figure~\ref{fig:so7'}.
Extending the search along the blue line of vanishing  $\phi'(0)$, we find that there are no other intersection with any red lines, i.e.\ no other solution to (\ref{eq:phiprime}) in the parameter space.

\subsubsection{$A\not=0$}

We now extend the search of solutions to the full truncation of G$_2$-invariant singlets,
i.e.\ we allow for non-vanishing $A$. In this case we scan the three-dimensional parameter space
(\ref{initial}) for regular solutions. Similar to our discussion of the SO(7) sector, we start by restricting the search
to even solutions
\begin{equation}
\phi(-w)=\phi(w)\,,\quad
\Delta(-w)=\Delta(w)\,,\quad
A(-w)=A(w)\,,
\label{eq:evenA}
\end{equation}
again corresponding to a $\mathbb{Z}_2$-symmetry of the system (\ref{systEQG2}). 
Accordingly, starting from a solution regular at $w=1$, the symmetry (\ref{eq:evenA}) is implemented
by the following conditions
\begin{equation}
\phi'(0)=0=\Delta'(0)=A'(0)
\;.
\label{condAeven}
\end{equation}
at $w=0$. Regularity at the other endpoint $w=-1$ is then implied by 
the symmetry (\ref{eq:evenA}). 
Consequently, we scan the three-dimensional parameter space for points where all
three conditions (\ref{condAeven}) hold exactly.
Generalising the analysis of section~\ref{subsec:A0},
we study the intersections of the hyperplanes, defined by the vanishing of
$\phi'(0)$, $\Delta'(0)$, and $A'(0)$, respectively.
However, this analysis reveals only the known solutions SO(7)$_-$ and G$_2$, 
listed in Table~\ref{tab:solutions} and (\ref{sol_analytic}).

\begin{figure}[bt]
\center
\includegraphics[scale=.70]{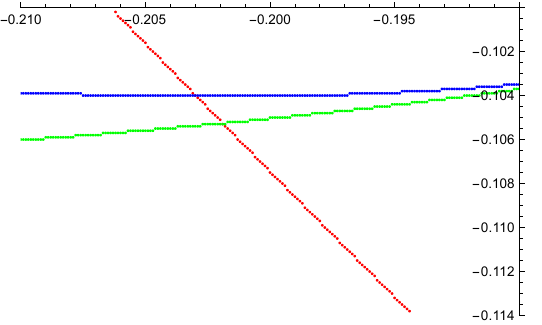}\qquad
\includegraphics[scale=.70]{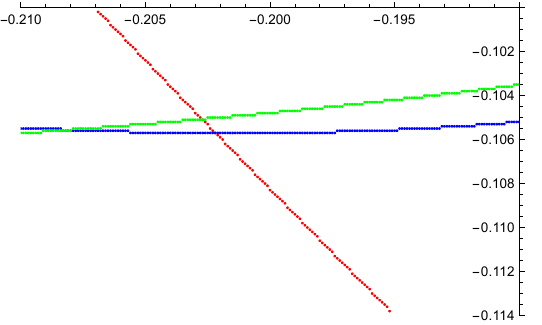}
\caption{Lines of vanishing $\phi'(0)$ (red), vanishing $\Delta'(0)$ (blue) and vanishing $A(0)$ (green)
on slices in the parameter space of initial conditions, with $\mathfrak{q}$ on the horizontal axis and $\mathfrak{p}$ on the vertical axis. 
The two slices are given at the values $\mathfrak{a}_1=-0.226667$ (left) and $\mathfrak{a}_2=-0.225417$, respectively.
The common intersection of red, blue and green line, which must appear on some intermediate slice, corresponds to the solution G$_2'$.
}
\label{fig:linesV}
\end{figure}
\begin{figure}[bt]
\center
\includegraphics[scale=.54]{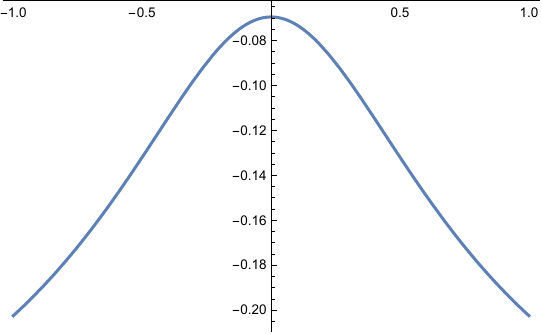}\quad
\includegraphics[scale=.54]{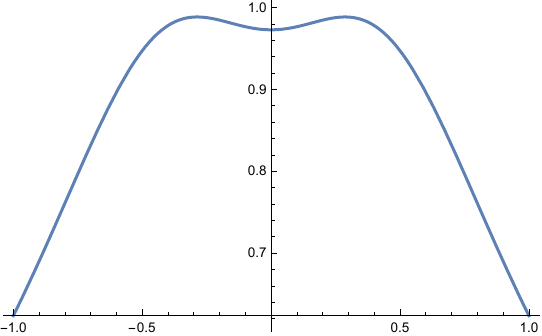}\quad
\includegraphics[scale=.54]{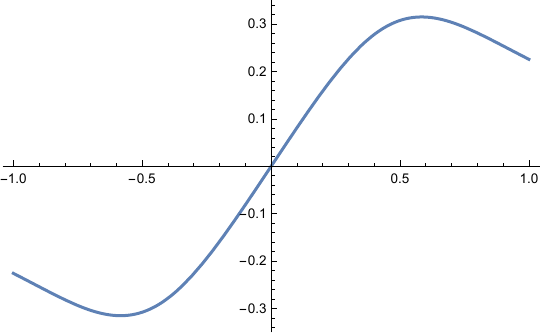}
\caption{New solution G$_2'$: fields $\phi$, $\Delta$, and $A$  as functions of $w$.}
\label{fig:g2'}
\end{figure}

Next, we employ another $\mathbb{Z}_2$-symmetry of the system (\ref{systEQG2}):
$A\rightarrow\pm A$,
and search for solutions in which $\phi$ and $\Delta$ are even whereas $A$ is odd in $w$
\begin{equation}
\phi(-w)=\phi(w)\,,\quad
\Delta(-w)=\Delta(w)\,,\quad
A(-w)=-A(w)\,.
\label{eq:oddA}
\end{equation}
Similar to (\ref{condAeven}), the symmetry (\ref{eq:oddA}) can be implemented
by the following conditions 
\bea
\phi'(0)=0=\Delta'(0)=A(0)
\;,
\label{condAodd}
\eea
at $w=0$, which in turn implies regularity throughout the interval.
We search for such solutions with the same method described above.
In Figure~\ref{fig:linesV}, we have depicted two slices in the three-dimensional parameter space,
defined by fixed neighboured values $\mathfrak{a}_1$, $\mathfrak{a}_2$, of $\mathfrak{a}$. 
In each slice we plot the three
curves defined by the vanishing of $\phi'(0)$, $\Delta'(0)$, and $A(0)$, respectively.
The configuration of the lines shows that on some intermediate slice 
$\mathfrak{a}_1<\mathfrak{a}<\mathfrak{a}_2$
there must be a common intersection point
of the three lines.
Having estalished its existence, we can then zoom in and optimise the numerical accuracy of the solution.
The result is the new solution called G$_2'$ in Table~\ref{tab:solutions}.
The corresponding profiles of the fields $\phi$, $\Delta$ and $A$ are plotted in Figure~\ref{fig:g2'}.

\begin{figure}[bt]
\center
\includegraphics[scale=.50]{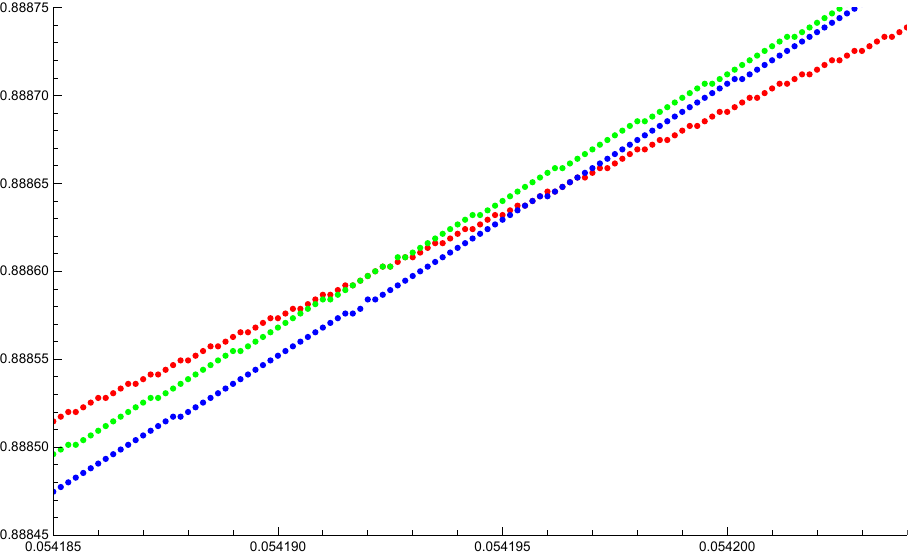}\;\;\;
\includegraphics[scale=.50]{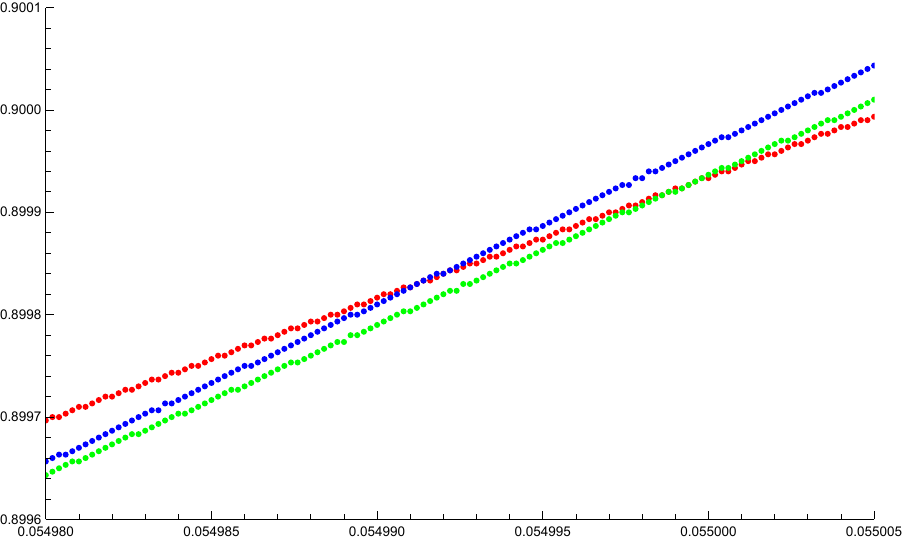}
\caption{Lines of vanishing $\phi'(0)$ (red), vanishing $\Delta'(0)$ (blue) and vanishing $A(0)$ (green)
on slices in the parameter space of initial conditions, with $\mathfrak{q}$ on the horizontal axis and $\mathfrak{p}$ on the vertical axis. 
The two slices are given at the values $\mathfrak{a}=0.658$ (left) and $\mathfrak{a}=0.66$, respectively.
The common intersection of red, blue and green line, which must appear on some intermediate slice, would correspond to the solution G$_2''$.
}
\label{fig:linesV2}
\end{figure}

\begin{figure}[bt]
\center
background\includegraphics[scale=.54]{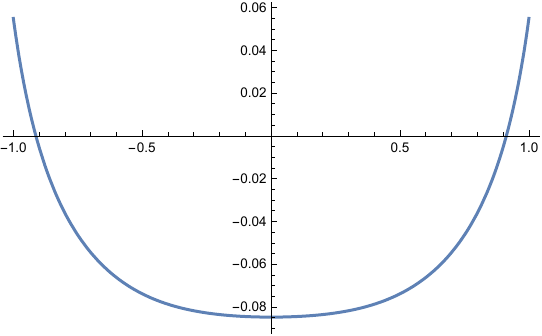}\quad
\includegraphics[scale=.54]{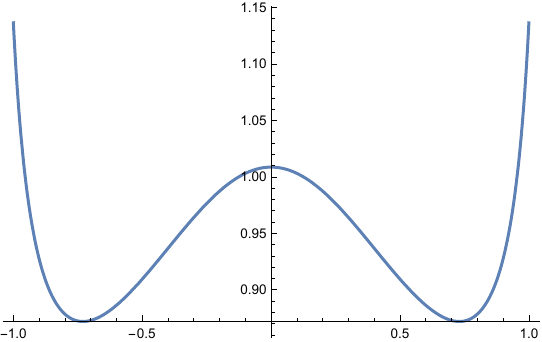}\quad
\includegraphics[scale=.54]{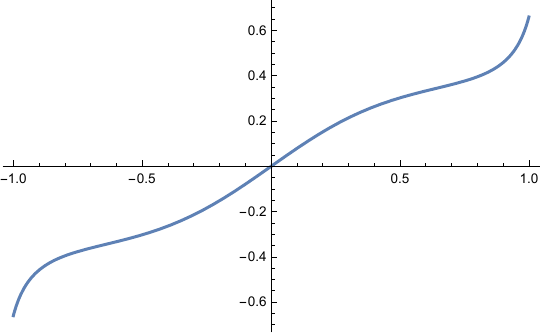}
\caption{New solution G$_2''$: fields $\phi$, $\Delta$, and $A$  as functions of $w$.}
\label{fig:g2''}
\end{figure}

It remains to extend the analysis to the full parameter space by systematically scanning 
the two-dimensional slices of fixed $\mathfrak{a}$.
Zooming into a different area in parameter space, we have also
identified the slices shown in Figure~\ref{fig:linesV2}.
Again, the configuration of lines indicates the existence of an intermediate slice
with a common intersection of all three lines, thus another exact solution
to (\ref{condAodd}).
The resulting solution is given as 
G$_2''$ in Table~\ref{tab:solutions}.
The corresponding profiles of the fields $\phi$, $\Delta$ and $A$ are plotted in Figure~\ref{fig:g2''}.

We have further gone through the slices of the three-dimensional parameter space and not
found any other critical region that would indicate another solution. Although we have not attempted
a rigorous proof, the analysis suggests that the set of regular solutions given in Table~\ref{tab:solutions}
is complete, if one restricts to even (\ref{condAeven}) and odd (\ref{condAodd}) solutions.
Relaxing the latter conditions, one may expect yet more regular solutions, but we have not explored this systematically.

\begin{figure}[bt]
\center
\includegraphics[scale=.59]{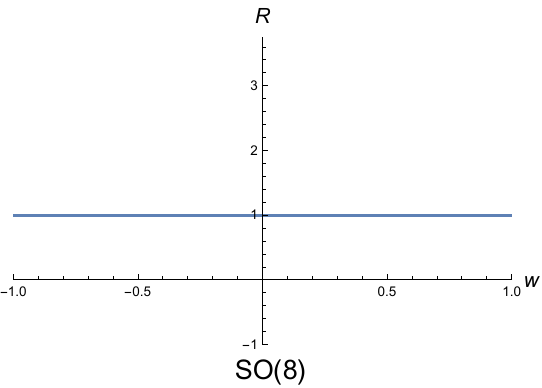}
\includegraphics[scale=.59]{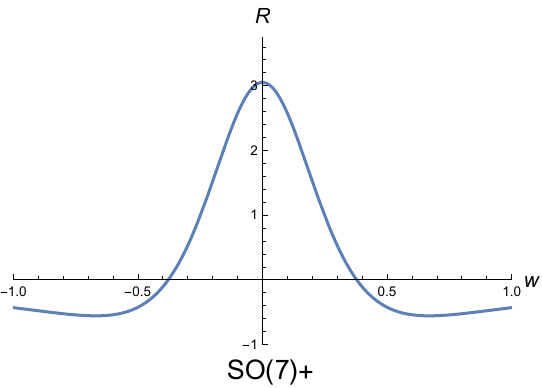}
\includegraphics[scale=.59]{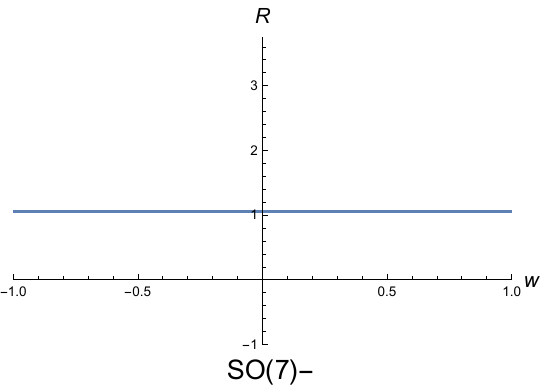}\\[2ex]
\includegraphics[scale=.59]{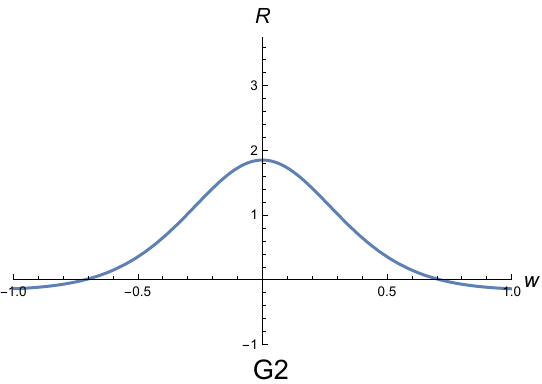}
\includegraphics[scale=.59]{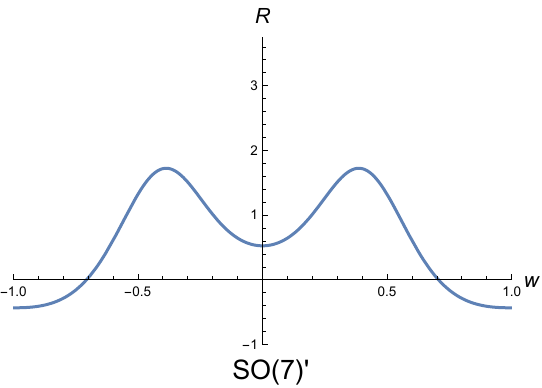}
\includegraphics[scale=.59]{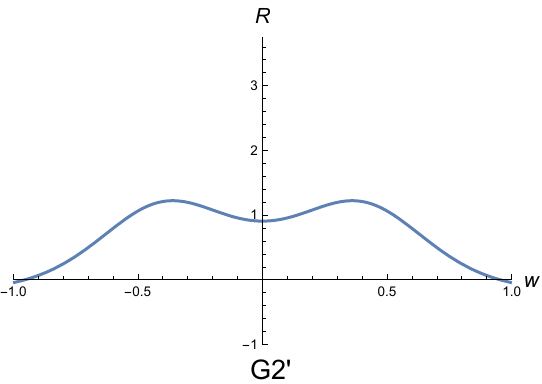}\\[2ex]
\includegraphics[scale=.59]{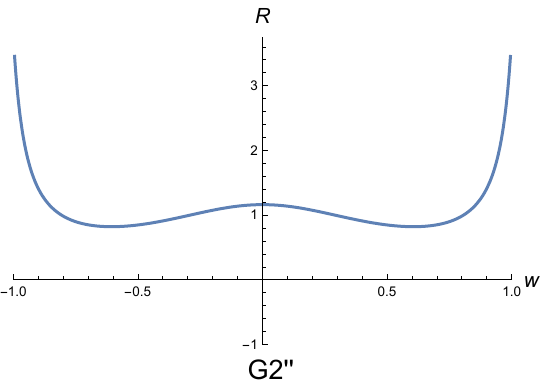}
\caption{
Curvature scalar $R_7$ of the internal deformed $S^7$ as a function of $w\in[-1,1]$ for the different 
G$_2$-invariant solutions collected in Table~\ref{tab:solutions}.
}
\label{fig:Rsolutions}
\end{figure}

Let us recall that the spacetime geometry for all of them 
is of the form AdS$_4\times \Sigma_7$, c.f.\ 
(\ref{metric11DG2-2}), where the internal space $\Sigma_7$ is given by 
a squashed seven sphere preserving ${\rm SO}(7)$ isometries.
In order to characterise the different geometries, we compute the curvature scalar
$R_7$ of the internal manifold $\Sigma_7$.
From (\ref{metric11DG2-2}), and using the equations of motion (\ref{systEQG2}) and (\ref{VD11G2})
to simplify the expression, we obtain the following expression
\bea
R_{7} &=&
6\, e^{-3 \phi/2}\,\Delta^{-1/2} \, \left(4 w  A-(1-w^2)\, \partial_w A\right)^2
+96 \,F^2 \,(1-w^2)\, e^{3 \phi/2}\, \Delta ^{5/2}  A^2
+72 \,e^{2 \phi }\,\Delta^{-2}\,A^2 
\nonumber\\
&&{}
+40 \,F^2\, \Delta^4
-\frac{12\, \Delta}{\ell_4^2}
-3 \,(1-w^2) \,e^{-3 \phi } \,\Delta^{-1}  (\partial_w\Delta)^2
\;,
\eea
in terms of the fields $\phi$, $\Delta$, and $A$.
As an illustration, we may plot the resulting function for the different solutions of Table~\ref{tab:solutions},
which is displayed in Figure~\ref{fig:Rsolutions}.

\subsection{Numerics}

In the previous section, we have established the existence of regular solutions at certain
discrete points in the three-dimensional parameter space.
For each solution, once we have proven its existence, we can zoom in to improve the numerical accuracy.
To this end, we have finally implemented a simple gradient descent algorithm in Python. 
As explained above, the problem is set by three initial conditions (\ref{initial}).
Next, we define a regularisation function, or loss function, to assess the regularity of a given solution. 
Put differently, this function quantifies how far a solution is from being regular. 

As discussed above, for even and odd solutions, 
regularity is conveniently encoded in the conditions (\ref{condAodd}) and (\ref{condAodd}), respectively.
Accordingly, we can define the loss function as
\begin{equation}
    \mathcal{L} = \mathrm{ln}(\phi'(0)^2 + \Delta'(0)^2). 
\end{equation}
when $\mathfrak{a} = 0$, and as
\bea
\mathcal{L} &=& \mathrm{ln}(\phi'(0)^2 + \Delta'(0)^2+A'(0)^2)\,,\nonumber\\
\mathcal{L} &=& \mathrm{ln}(\phi'(0)^2 + \Delta'(0)^2+A(0)^2)\,.
\eea
for even (\ref{eq:evenA}) and odd (\ref{eq:oddA}) solutions, respectively.

With these loss functions in place, we can perform gradient descent by updating the initial parameters according to
\begin{equation}(\mathfrak{p},\mathfrak{q},\mathfrak{a}) \longleftarrow (\mathfrak{p},\mathfrak{q},\mathfrak{a}) - \alpha \nabla \mathcal{L}(\mathfrak{p},\mathfrak{q},\mathfrak{a})
\end{equation}
where $\alpha$ is the learning rate, controlling the step size in the gradient descent. This method allows us to verify and refine the previous analysis, enabling a fine-tuning of the initial parameters. The results are collected in Table~\ref{tab:solutions} where all numbers are accurate to the displayed digits.

\section{Conclusions}
\label{sec:conclusion}

In this paper, we have discussed the consistent truncations to ${\rm K}$-singlets w.r.t.\
to a subgroup ${\rm K}$ of the isometry group of the internal manifold.
We have reviewed how these truncations are described in the framework of generalised geometry
and exceptional field theory.
As an application, we have worked out the field equations for the most general 
G$_2$-invariant AdS$_4$ solution of ${D=11}$ supergravity, with the internal space $\Sigma_7$ given by
a squashed seven-sphere preserving SO(7) isometries.
The ExFT description of this truncation features a scalar sector described by the six-dimensional 
coset space 
$\left({{\rm SU}(2,1)}\times{{\rm SU}(1,1)}\right) / \left( {{\rm U}(2)}  \times{{\rm U}(1)}\right)$
with all scalars still depending on an extra coordinate~$\theta$. The latter encodes the description of
the infinite Kaluza-Klein towers of G$_2$-singlets within a four-dimensional field theory.
Searching for AdS$_4$ vacua, we have shown that the system can be simplified to a set
of three second-order ordinary differential equations for three scalar fields. Furthermore, we have given the explicit uplift 
of this sector to $D=11$ dimensions.

Imposing a compact internal seven-dimensional space restricts the search to solutions regular at the endpoints of the interval $\theta\in[0, \pi]={\cal I}$,
with the seven-sphere represented as a foliation of $S^6 = {{\rm G}_2}/{{\rm SU}(3)}$ over the interval ${\cal I}$\,.
The equations of motion are singular at these endpoints and a closer inspection shows that only a discrete set of such regular solutions exists.
More precisely, solutions that are regular at one endpoint $\theta=0$ are characterised  by a three-dimensional parameter space. Imposing regularity throughout the interval defines discrete points in this space. We conduct a numerical scan for these points. 
Importantly, we find that the condition of regularity can be very efficiently
implemented by requiring the solutions to be even/odd according to (\ref{eq:evenA}) or (\ref{eq:oddA}),
such that regularity at the opposite endpoint follows from symmetry.
In this sector, we recover in particular the four solutions that were previously known in analytic form \cite{deWit:1984nz}. These all 
live within the consistent truncation to ${\cal N}=8$ supergravity \cite{deWit:1982bul} and correspond to the 
four G$_2$-invariant extremal points of its scalar potential \cite{Warner:1983vz}. 
On top of these known solutions, we identify three new numerical regular solutions,
which we label as ${\rm SO}(7)'$, ${\rm G}_2'$, and ${\rm G}_2''$, respectively.
They all uplift to $D=11$ geometries of the form AdS$_4\times \Sigma_7$
together with a non-vanishing three-form flux which preserves G$_2 \subset {\rm SO}(7)$ symmetry.
All these solutions are collected in Table~\ref{tab:solutions}.
Within the sector of even/odd solutions satisfying (\ref{eq:evenA}) or (\ref{eq:oddA}), the analysis appears to be complete.
Relaxing these additional conditions, one may expect yet more regular solutions, 
and it would be highly interesting to extend the numerical search to be able to identify all the regular solutions of the system.

The embedding of the new solutions into the ExFT framework allows to directly extract the generalised frames associated to these backgrounds. In turn, that should allow to adapt the techniques of \cite{Malek:2019eaz,Malek:2020yue,Cesaro:2020soq,Duboeuf:2022mam} for a computation of the Kaluza-Klein spectra around these new backgrounds. 
It would be particularly interesting to find if supersymmetry is preserved by any of these backgrounds.

Remarkably, most of the solutions we have identified already live within the consistent truncation to ${\cal N}=8$ supergravity. 
I.e.\ they only require non-vanishing scalar fields from the lowest Kaluza-Klein multiplet. Allowing for non-vanishing scalars among 
the infinitely many higher Kaluza-Klein modes somewhat surprisingly only gives rise to three new AdS$_4$ solutions in this sector.
In turn, it is then tempting to speculate that these new solutions might also be related to some particular consistent 
truncations of the full theory.
It may be worth noting that the $\omega$-deformed maximal 
supergravities constructed in \cite{DallAgata:2012bb} do 
admit additional G$_2$-invariant vacua while retaining the
${\cal N}=8$ vacuum of the round sphere \cite{DallAgata:2012bb,Borghese:2012qm,Berman:2022jqn}. 
Yet, it probably is wishful musing to imagine that these theories
 might play a role in the description of the new vacua.
While that would certainly be an exceptional turn of events, 
we leave these questions and others for future studies.

\subsection*{Acknowledgements}
We would like to thank Franz Ciceri, Camille Eloy, Hermann Nicolai and Ergin Sezgin for stimulating discussions. The work of MG has been supported by the Australian Research Council (ARC) Future Fellowship FT180100353, the ARC Discovery Project DP240101409, and by a Capacity Building Package of the University of Queensland.

\providecommand{\href}[2]{#2}\begingroup\raggedright\endgroup

\end{document}